\documentstyle[a4,amstex,12pt]{article}
\newcommand{\real}{{\Bbb R}}
\newcommand{\cplx}{{\Bbb C}}
\newcommand{\zint}{{\Bbb Z}}
\newcommand{\cplxn}{\cplx^{n}}
\newcommand{\Ncplxn}{\otimes^N \cplx^{n} }
\newcommand{\cz}{\cplx[z_1^{\pm 1},z_2^{\pm 1},\dots,z_N^{\pm 1}]}
\newcommand{\hs}[1]{{\cal H}^{(#1)}}
\newcommand{\h}{{\cal H}}
\newcommand{\ep}{\epsilon}
\newcommand{\gl}{ {{\frak g} {\frak l}}}
\newcommand{\gln}{ {\frak g}{\frak l}_n}
\newcommand{\Sgroup}[1]{{\frak S}_{#1}}
\newcommand{\id}{ {\mathrm i}{\mathrm d}}
\newcommand{\Ale}{{\frak A}}
\newcommand{\p}{\partial}
\newcommand{\sprod}[2]{\langle #1, #2 \rangle}
\newcommand{\qdet}{{\mathrm q}\det}
\newcommand{\s}{\sigma}
\renewcommand{\l}{\lambda}
\newcommand{\la}{\lambda}
\renewcommand{\o}{\omega}
\newcommand{\mm}{{\bold m}}
\newcommand{\mlett}{m}
\newcommand{\nlett}{n}
\newcommand{\nn}{{\bold n}}
\newcommand{\MC}{{\cal M}}
\newcommand{\KKi}{K_{i,i+1}}

\newcommand{\halmos}{\rule{5pt}{5pt}}
\newcommand{\ba}{\begin{array}}
\newcommand{\ea}{\end{array}}
\newcommand{\beq}{\begin{equation}}
\newcommand{\eeq}{\end{equation}}
\newcommand{\bqa}{\begin{eqnarray}}
\newcommand{\eqa}{\end{eqnarray}}
\newcommand{\bqas}{\begin{eqnarray*}}
\newcommand{\eqas}{\end{eqnarray*}}

\numberwithin{equation}{section}

\newtheorem{prop}{\bf Proposition}
\newtheorem{thm}{\bf Theorem}
\newtheorem{lemma}{\bf Lemma}

\newenvironment{pf}{\noindent {\it {\normalsize P}\normalsize{roof.}}  
\normalsize\hskip 5pt}{\hfill\halmos}

\begin{document}

\begin{titlepage}
\pagestyle{empty}
\flushright{RIMS-1114}
\begin{center}
\mbox{} \\
\vspace{2.7cm}
\begin{Large}
{\bf The orthogonal eigenbasis and norms of eigenvectors in  the Spin Calogero-Sutherland Model } 
\end{Large} \\
\vspace{1.5cm}
\large{ Kouichi Takemura }\footnote{e-mail: takemura@@kurims.kyoto-u.ac.jp} 
\large{ and Denis Uglov} \footnote{e-mail: duglov@@kurims.kyoto-u.ac.jp}  
\\  Research Institute for Mathematical Sciences\\
Kyoto University, Kyoto 606, Japan  \\ 
\vspace{1cm}
November 1996  \\

\vspace{2cm}

\begin{abstract}
Using a technique based on the Yangian Gelfand-Zetlin algebra and the associated Yangian Gelfand-Zetlin bases we construct an orthogonal basis of eigenvectors in the Calogero-Sutherland Model with spin, and derive product-type formulas for norms of these eigenvectors.

\end{abstract}

\end{center}
\end{titlepage}

\section{Introduction}
In this paper we study the spin generalization of the Calogero-Sutherland Model which was proposed in \cite{BGHP}. This Model describes $N$ particles with coordinates $x_1, x_2,\dots,x_N$ moving along the circle of the unit radius $( 0 \leq x_i \leq 2\pi )$. Each particle carries a spin with $n$ possible values, and the dynamics of the Model are governed by the Hamiltonian  
\begin{equation}
H_{SCSM} = - \sum_{i=1}^N \frac{\p^2}{\p x_i^2} + \frac{1}{4} \sum_{1\leq i \neq j \leq N} \frac{ \beta \left( P_{i,j} + \beta \right)}{ \sin^2\left(\frac{x_i - x_j}{2}\right) } \label{eq:H1}
\end{equation}
where $ \beta$ is a coupling constant and the $P_{i,j}$ is the spin exchange operator for the particles $i$ and $j$.

The scalar version of the $H_{SCSM}$ $(n=1)$  has been studied over the course of the past 25 years starting with the work of Sutherland \cite{Sutherland}. Among the recent advances one can point out the connection of the $H_{SCSM}$ $(n=1)$ with Random Matrix Theory \cite{Forresterrandom}, exact computation of the dynamical correlation functions \cite{Ha, Lesage, Minahan} and the intriguing connection with the Virasoro and the $W$-algebras \cite{Awata}. To a large extent many of these developments, in particular the computation of the correlation functions, were based on the properties of the symmetric Jack polynomials which describe the orthogonal eigenbasis of the scalar Calogero-Sutherland Model \cite{Stanley, Macdonaldbook}. 

Considerably less is known about the Calogero-Sutherland Model with spin $(n \geq 2)$. In the work \cite{BGHP} the construction of eigenvectors for the Calogero-Sutherland Model with general spin has been proposed. This construction is based on the diagonalization of the Dunkl operators \cite{Dunkl} by the non-symmetric Jack polynomials. Although the way to obtain the eigenvectors was pointed out in \cite{BGHP}, the complete and orthogonal eigenbasis has not been constructed so far.

In the present paper we give a construction of such an eigenbasis in terms of the non-Symmetric Jack polynomials and derive explicit product-type formulas for the norms of the eigenvectors. 

In the case of the scalar Model the knowledge of explicit formulas for the norms of the Jack polynomials has been  essential for the computation of the dynamical correlation functions. Therefore we believe that  the results of our present work will  turn out to be of use in the computation of the 2-point dynamical correlation functions in the Spin Calogero-Sutherland Model.   

Let us now describe the main features of our construction. The principal role in it is played by the Yangian symmetry of the Spin Calogero-Sutherland Model. As was discovered and emphasized in \cite{BGHP}, the space of states in the Model admits the action of the algebra $Y(\gln)$ -- the Yangian of $\gln$ \cite{Drinfeld}, \cite{nt2}. This action is given by the $n\times n$ operator-valued monodromy matrix $ \| T_{a,b}(u) \|_{1\leq a,b \leq n} $ which is regarded as the formal Taylor series in negative powers of the spectral parameter $u$. The center of the Yangian action is generated by the operator coefficients $\Delta^{(s)}$ in the expansion of the quantum determinant ${\qdet}T(u)$ of the monodromy matrix:  
\begin{gather}
{\qdet}T(u) = \sum_{\s \in \Sgroup{n}} (-1)^{l(\s)} T_{1,\s(1)}(u)T_{2,\s(2)}(u-1)\cdots T_{n,\s(n)}(u-n+1) = \sum_{s=0}^{\infty} u^{-s}\Delta^{(s)}, \\
\mbox{} [ T_{a,b}(u), \Delta^{(s)}] = 0 \qquad ( a,b =1,2,\dots,n;\quad s=0,1,2,\dots\;).
\end{gather}
The Hamiltonian of the Model belongs to the Abelian algebra generated by the conserved charges $\Delta^{(s)}$ \cite{BGHP} and thereby commutes with the Yangian action. 

In the scalar case $(n=1)$ the Yangian $Y({\frak g}{\frak l}_1)$ coincides with its center and is just the algebra of the conserved charges in the Calogero-Sutherland Model. It is known \cite{MacdonaldLN, Macdonaldbook} that in this case the joint spectrum of the conserved charges is simple, and that the operators   $\Delta^{(s)}$ are self-adjoint with respect to the scalar product relevant for the computation of quantities such as correlation functions. Hence the orthogonal eigenbasis of $H_{SCSM}$ $(n=1)$ is defined uniquely up to normalizations of eigenvectors as the eigenbasis of the Abelian algebra generated by the conserved charges $\Delta^{(s)}$.

In the situation when spin is non-trivial $(n \geq 2)$ the spectrum of the quantum determinant is not simple  and thus the higher conserved charges alone are not sufficient to specify an orthogonal eigenbasis. To give such a specification, in this paper we use a maximal Abelian sub-algebra of $Y(\gln)$ denoted by $A(\gln)$ and known as the Yangian  Gelfand-Zetlin algebra. This algebra includes the center of the Yangian as a sub-algebra. The algebra $A(\gln)$ was first studied  by Cherednik \cite{C} and subsequently by Nazarov and Tarasov \cite{nt1,nt2}. It is defined as the sub-algebra in $Y(\gln)$ generated by  all the centers in the chain of algebras  
\begin{equation}
Y({\frak g}{\frak l}_1) \subset Y({\frak g}{\frak l}_2) \subset \cdots \subset Y(\gln)
\end{equation}
where $Y({\frak g}{\frak l}_{m-1})$ is realized inside $Y({\frak g}{\frak l}_m)$ as the sub-algebra generated by the entries of the sub-matrix  $ \| T_{a,b}(u) \|_{1\leq a,b \leq m-1} $.

The generators of the Abelian algebra $A(\gln)$ which appear in the Spin Calogero-Sutherland Model possess the two crucial properties: 
\begin{align*}
 & \bullet \; \text{They are self-adjoint with respect to the relevant scalar product (defined in sec. \ref{sec:definition}).} \\ 
 & \bullet \; \text{They are simultaneously diagonalizable and their joint spectrum is simple. }
\end{align*}
 \mbox{} From these two properties it follows that,  since the Spin Calogero-Sutherland Hamiltonian belongs to the algebra  $A(\gln)$, the eigenbasis of the algebra $A(\gln)$ is an orthogonal eigenbasis of the  Hamiltonian. 

Construction of this eigenbasis is the first main problem which we address in the present paper. This  construction is carried out in two steps. First, we describe the decomposition of the space of states in the  Model into irreducible sub-representations of the Yangian action and point out the  Yangian highest-weight vector in each of the irreducible components. These highest-weight vectors are expressed in terms of the non-symmetric Jack polynomials. 

In the paper \cite{nt2} Nazarov and Tarasov gave  construction of canonical bases, called Yangian Gelfand-Zetlin bases, for  a wide class of Yangian representations which includes all representations which appear as irreducible components of the Yangian action in the Spin Calogero-Sutherland Model. The Yangian Gelfand-Zetlin base of Nazarov and Tarasov is defined as the base where the action of the Abelian algebra $A(\gln)$ is diagonal. It includes the highest-weight vector and ``descendants'' which are obtained by acting on the highest-weight vector with appropriate creation operators described explicitly in \cite{nt2}.   

Once we have found the irreducible Yangian decomposition of the space of states and have identified the highest-weight vectors, the results of \cite{nt2} can immediately be applied  to describe the eigenbasis of $A(\gln)$ within each of the irreducible sub-representations and hence in the entire space of states of the Model. 

The second main problem which we consider in this paper is computation of the norms of the eigenvectors. This computation is performed as follows. First, the norms of the Yangian highest-weight vectors are found by expressing them in terms of  the norms of the non-symmetric Jack polynomials known from \cite{Cherednik2, Macdonald1, Opdam}, then the norms of the ``descendants'' are computed by using properties of the creation operators.  

Now let us describe the contents of the paper. In  sec. \ref{sec:definition} we recall the definition of the Spin Calogero-Sutherland Model. In sec. \ref{sec:background} the necessary background information on the Yangian and Yangian Gelfand-Zetlin bases is reviewed. The contents of this section largely follow the work \cite{nt2}. In sec. \ref{sec:yangian} we discuss properties of Yangian action in the Spin Calogero-Sutherland Model. In sec. \ref{sec:decomposition} the irreducible Yangian decomposition of the space of states is given. The main results in this section are theorems \ref{t:FDEC} and \ref{t:BDEC}. Section \ref{sec:hwvnorms} contains derivation of the norm formulas for the Yangian highest-weight vectors. The main result here is the proposition \ref{p:prop4}. In sec. \ref{sec:dnorms} we give expressions for the ``descendants'' of the highest-weight vectors. The proposition \ref{norm} gives  formulas for their norms. 

The Appendix contains proofs of some of the statements in the main text.

\section{Definition of the Model} \label{sec:definition}
In this section we will review the definition and a few known facts about the Spin Calogero-Sutherland Model (SCSM). In doing so we will closely follow the work \cite{BGHP} where this Model was introduced and extensively studied for the first time under the name of the Dynamical Model with long-range interaction. We would like to note, that the Model which we define below is the gauge transformed version of (\ref{eq:H1}) \cite{BGHP}. 

\subsection{The Hilbert space of states in the gauge transformed SCSM}
The space of states of the gauge transformed SCSM \cite{BGHP} is a subspace in the tensor product
\beq
{\cal H} := \cz\otimes(\Ncplxn).
\end{equation}
We fix the base $\{v_{\ep}\}_{\ep=1,\dots,n}$ in $\cplxn$ and define in $\Ncplxn$ the hermitian (sesquilinear) scalar product ${\sprod{\;\cdot\;}{\;\cdot\;}}_s$ by requiring pure tensors to be orthonormal: 
\beq
{\sprod{v_{\ep_1}\otimes v_{\ep_2}\otimes \cdots \otimes v_{\ep_N}}{v_{\tau_1}\otimes v_{\tau_2}\otimes \cdots \otimes v_{\tau_N}}}_s:= \prod_{i=1}^N \delta_{\ep_i,\tau_i}\qquad (\ep_i,\tau_i = 1,2,\dots,n). \label{eq:sps}
\end{equation}
In $\cz$ we define the hermitian scalar  product ${\sprod{\;\cdot\;}{\;\cdot\;}}_c$ which depends on the parameter $\alpha \in \real_{>0}$. For $f(z_1,z_2,\dots,z_N)$, $g(z_1,z_2,\dots,z_N)$ $\in$ $\cz$ set
\begin{equation}  
{\sprod{f}{g}}_c := \frac{1}{N!}\left( \prod_{i=1}^N \oint_{|w_i|=1} \frac{dw_i}{2\pi \sqrt{-1}w_i}  \right) \left( \prod_{ i\neq j} 1 - \frac{w_i}{w_j} \right)^{\frac{1}{\alpha}} \overline{f(w_1,w_2,\dots,w_N)} g(w_1,w_2,\dots,w_N) \label{eq:spc} 
\end{equation}
where the integration  over each of the complex variables $w_i$ is taken along the unit circle in the complex plane. The hermitian scalar product ${\sprod{\;\cdot\;}{\;\cdot\;}}$ in the space $\h$ is defined as the composition of the scalar products (\ref{eq:sps}) and (\ref{eq:spc}). For $f, g$ $\in$ $\cz$ and $u, v$ $\in$ $\Ncplxn$ put   
\begin{equation}
\sprod{f\otimes u}{g\otimes v} := {\sprod{f}{g}}_c{\sprod{u}{v}}_s \label{eq:sp}
\end{equation}
and extend the ${\sprod{\;\cdot\;}{\;\cdot\;}}$ on the entire space $\h$ by requiring it to be sesquilinear.

The symmetric group $\Sgroup{N}$ acts in the $\h$. For $\s =$ \begin{small}$\left(\ba{cccc} 1 & 2 & \cdots & N \\ \s(1) & \s(2) & \cdots & \s(N) \ea \right)$ \end{small} $\in \Sgroup{N}$ there are two  right actions $K_{\s}$ and $P_{\s}$ defined in the base $\{ z_1^{m_1}z_2^{m_2}\cdots z_N^{m_N}\otimes v_{\ep_1}\otimes v_{\ep_2}\otimes \cdots \otimes v_{\ep_N} \}$ $(m_i\in \zint,\; 1 \leq \ep_i \leq n)$ of the space $\h$ by  
\begin{equation}
\left.\ba{c} K_{\s} \\ P_{\s} \ea  \right\} \cdot z_1^{m_1}\cdots z_N^{m_N}\otimes v_{\ep_1}\otimes \cdots \otimes v_{\ep_N} = \begin{cases} z_1^{m_{\s(1)}}\cdots z_N^{m_{\s(N)}}\otimes v_{\ep_1}\otimes \cdots \otimes v_{\ep_N}, \\ z_1^{m_1}\cdots z_N^{m_N}\otimes v_{\ep_{\s(1)}}\otimes  \cdots \otimes v_{\ep_{\s(N)}}. \end{cases}
\end{equation}
For the transposition $(i,j)\in \Sgroup{N}$ we will use the notations  
\begin{equation}
K_{(i,j)} \equiv K_{i,j}\quad  \text{and} \quad P_{(i,j)} \equiv P_{i,j}.
\end{equation}
The operators $K_{i,j} $ and $P_{i,j}$ are easily seen to be self-adjoint and unitary with respect to the scalar product (\ref{eq:sp}).

The SCSM can be defined in two versions -- fermionic and bosonic \cite{BGHP}. Throughout this paper we will distinguish these versions by the sign of the parameter $\kappa$ setting $ \kappa = - $ (resp. $ \kappa = + $) for the fermionic (resp. bosonic) case. The space of states $\hs{\kappa}$ in the gauge transformed SCSM is then defined as follows: 
\begin{equation}
\hs{\kappa}:= \bigcap_{i=1}^{N-1} Ker(K_{i,i+1}P_{i,i+1} - \kappa 1) \subset \h. \label{eq:Ker}
\end{equation}
Or, equivalently, as the image of the total antisymmetrization or symmetrization  operator: 
\begin{equation}
\Ale_N^{(\kappa)} := \sum_{\s \in \Sgroup{N}} (\kappa)^{l(\s)} K_{\s}P_{\s}. \label{eq:anti} 
\end{equation}
The subspace $\hs{\kappa}$ inherits the scalar product $(\ref{eq:sp})$ from the space $\h$. We will use the notation ${\sprod{\;\cdot\;}{\;\cdot\;}}_{(\kappa)}$ for this scalar product.

\subsection{The Hamiltonian of the SCSM}
The gauge-transformed SCSM Hamiltonian is defined through the Dunkl operators \cite{Dunkl,BGHP}: 
\begin{equation}
d_i := \alpha\, z_i\frac{\p}{\p z_i} - i + \sum_{i < j}\frac{z_j}{z_j-z_i}(K_{i,j}-1) - \sum_{i > j}\frac{z_i}{z_i-z_j}(K_{i,j}-1) \qquad (i=1,2,\dots,N), \label{eq:Dunkl}
\end{equation}
which satisfy the relations of the degenerate affine Hecke algebra:
\begin{eqnarray} 
K_{i,i+1}d_i - d_{i+1}\KKi & = & 1, \label{eq:Hecke1}\\ 
\mbox{} [d_j, \KKi] & = & 0 \qquad (j\neq i,i+1), \\
\mbox{} [d_i, d_j] & = &  0.  \label{eq:Hecke3}
\end{eqnarray}
We will consider the Dunkl operators as acting either in $\cz$ or in the first factor in $\h = \cz \otimes (\Ncplxn)$ by expressions (\ref{eq:Dunkl}) and trivially in the second factor: $(\Ncplxn)$ without always giving exact specification since this is unlikely to cause any confusion.

The relations (\ref{eq:Hecke1} - \ref{eq:Hecke3}) imply, in particular, that symmetric polynomials in $d_1,d_2,\dots, d_N$ leave the subspaces $\hs{\pm}$ invariant \cite{BGHP}.
In terms of the Dunkl operators the gauge-transformed Hamiltonian $H^{(\kappa)} \in End(\hs{\kappa})$ of the SCSM is   
\begin{multline}
 H^{(\kappa)} :=   \sum_{i=1}^N \left\{\left(\alpha\, z_i\frac{\p}{\p z_i}\right)^2 + (2i - N -1)\alpha\, z_i\frac{\p}{\p z_i}\right\} +  \label{eq:H}\\ 
  + 2\alpha \,\sum_{i<j}\left\{\frac{z_i}{z_i-z_j}\left(z_i\frac{\p}{\p z_i}-z_j\frac{\p}{\p z_j}\right) +  \frac{z_i z_j}{(z_i-z_j)(z_j-z_i)}(-\kappa P_{i,j}+1)\right\} + \frac{1}{12}N(N^2-1) =  \\
     =\sum_{i=1}^N \left(d_i - \frac{N+1}{2}\right)^2   - N(N+1)^2. \\ \mbox{}
\end{multline}
Here to show the equality one has to use the relation
\begin{equation} 
K_{i,j}f = \kappa P_{i,j}f  
\end{equation}
which holds for any $f\in \hs{\kappa}$ due to the definition (\ref{eq:Ker}). 

By a straightforward calculation one checks that the Dunkl operators are self-adjoint with respect to the scalar product (\ref{eq:spc}) and hence the Hamiltonian $H^{(\kappa)} $ is  self-adjoint with respect to the scalar product ${\sprod{\;\cdot\;}{\;\cdot\;}}_{(\kappa)}$. The physical Hamiltonian $H^{(\kappa)}_{SCSM}$ is obtained from the $H^{(\kappa)}$ by performing the gauge transformation \cite{BGHP}:  
\begin{equation}
H^{(\kappa)}_{SCSM} = V^{\frac{1}{\alpha}}H^{(\kappa)} V^{-\frac{1}{\alpha}} = \sum_{i=1}^{N}\left(\alpha\, z_i\frac{\p}{\p z_i}\right)^2 + \sum_{i\neq j}\frac{z_i z_j}{(z_i-z_j)(z_j-z_i)}(-\kappa\alpha P_{i,j}+1)
\end{equation}
where 
\begin{equation}
V = \left(\prod_{i=1}^N z_i\right)^{\frac{1-N}{2}}\prod_{i<j}z_i-z_j .
\end{equation}
The Hamiltonian $H^{(\kappa)}_{SCSM}$ is identified up to the overall factor $ \alpha^2$ with the Hamiltonian (\ref{eq:H1}) where $\beta = -\kappa 1/\alpha $ and $ z_i = \exp(\sqrt{-1}x_i)$. The $H^{(\kappa)}_{SCSM}$ is self-adjoint with respect to the physical scalar product which is obtained from (\ref{eq:sp}) by formally putting $\alpha = \infty$.


\section{Yangian $Y(\gln )$ and the Yangian Gelfand-Zetlin bases } \label{sec:background}
In this section we summarize properties of the Yangian $Y(\gln )$ which are used in this paper. The main attention is given to the Gelfand-Zetlin algebra and the canonical Yangian Gelfand-Zetlin bases in certain irreducible Yangian modules. The contents of this section, with the exception of the lemma \ref{lemma:anti-involution} can be found in the works \cite{nt1,nt2}. 

\subsection{The definition of the Yangian $Y(\gln )$
 and the Gelfand-Zetlin algebra} \label{gzalg}

The Yangian $Y(\gln )$ is a unital associative algebra generated
by the elements $1$ and $T_{a,b}^{(s)}$ where $a,b =1,\cdots,n$ and $s=1,2,\cdots$
that are subject to the following relations: 
\begin{equation}
[T_{a,b}^{(r)},T_{c,d}^{(s+1)}] - [T_{a,b}^{(r+1)},T_{c,d}^{(s)}] = T_{c,b}^{(r)}T_{a,d}^{(s)} - T_{c,b}^{(s)}T_{a,d}^{(r)} \qquad ( r,s = 0,1,2,\dots \;) 
\label{eq:Yangianrel}\end{equation}
where  $ T_{a,b}^{(0)} := \delta_{a,b}1$. 

Introducing the formal Taylor series in
$u^{-1}$
\begin{equation}
T_{a,b}(u)=\delta _{a,b} +T_{a,b}^{(1)}u^{-1}
+T_{a,b}^{(2)} u^{-2}+\ldots
\end{equation}
define $\stackrel{k}{T}(u) \: (k=1,2)$ as follows.
\begin{equation}
\stackrel{k}{T}(u)=\sum_{a,b=1}^{n} E_{a,b}^{(k)} \otimes T_{a,b}(u)
 \in End(\cplxn ) \otimes End(\cplxn ) \otimes Y(\gln )[[u^{-1}]].
\end{equation}
Here $E_{a,b}^{(k)}$ are the standard matrix units that are acting in the $k$-th tensor factor $\cplxn$.
If we put 
\begin{equation}
R(u,v)= id + \frac{1}{u-v} \sum_{a,b=1}^{n}  E_{a,b}^{(1)} 
\otimes E_{b,a}^{(2)}
\end{equation}
then the defining relations of $Y(\gln )$ are
\begin{equation}
R(u,v) \stackrel{1}{T} (u) \stackrel{2}{T}(v)
 = \stackrel{2}{T}(v) \stackrel{1}{T}(u) R(u,v). \label{RTT}
\end{equation}

 Let
${\bold i}=(i_1,\dots,i_m)$ and ${\bold j}=(j_1,\dots,j_m)$ be two
sequences of indices such that
\begin{equation}
1\leq i_1<\ldots<i_m\leq n
\; \mbox{ and } \;
1\leq j_1<\ldots<j_m\leq n.
\end{equation}
Let ${{\frak S} _{m}}$ be the symmetric group of degree $m$. Define  
\begin{equation}
Q_{{\bold i}{\bold j}}(u)=
\sum_{\sigma \in {{\frak S} _{m}}}
(-1)^{l(\sigma )}\cdot T_{i_1, j_{\sigma (1)}}(u) T_{i_2, j_{\sigma (2)}}(u-1)
\dots T_{i_m, j_{\sigma (m)}}(u-m+1),
\end{equation}
and
\begin{equation}
A_{0}(u)=1, \; \; \; A_{m}(u)= Q_{{\bold i}{\bold i}}(u) , \; \; (m=1, \cdots ,n)
\end{equation}
\begin{equation}
B_{m}(u)= Q_{{\bold i}{\bold j}}(u), \; \; 
C_{m}(u)= Q_{{\bold j}{\bold i}}(u), \; \; 
D_{m}(u)= Q_{{\bold j}{\bold j}}(u), \; \; ( m=1, \cdots ,n-1)
\end{equation}
 where ${\bold i}=(1,\ldots,m)$ and ${\bold j}=(1,\ldots,m-1,m+1)$.

 The following propositions are can be found in the paper \cite{nt1}.
\begin{prop} \label{pr1} {\em\cite{nt1} }
a) The coefficients of $A_n(u)$ belong to the center of the algebra
$Y(\gln )$. \\
 b) All the coefficients of $A_1(u),\dots,A_n(u)$ pairwise commute.
\end{prop}
\begin{prop} \label{pr2}  {\em\cite{nt1} }
The following commutation relations hold in $Y(\gln )$:
\begin{eqnarray}
 & [A_m(u),B_l(v)]=0 \; \; \mbox{if} \; \; \; \; l\neq m, & \\
 & [C_m(u),B_l(v)]=0  \; \; \mbox{if}  \; \; \; \;l\neq m, & \\
 & [B_m(u),B_l(v)]=0  \; \; \mbox{if}  \; \; \; \; |l-m|\neq1, & \label{BB} \\
 & (u-v) \cdot [A_m(u),B_m(v)]=B_m(u) A_m(v)-B_m(v) A_m(u), & \\
 & (u-v) \cdot [C_m(u),B_m(v)]=D_m(u) A_m(v)-D_m(v) A_m(u). & \label{CB}
\end{eqnarray}
\end{prop}
\begin{prop}{\em\cite{nt1} }
The following relation holds in $Y(\gln )$:
\begin{equation}
C_m(u) B_m(u-1)=D_m(u) A_m(u-1)-A_{m+1}(u) A_{m-1}(u-1). \label{CBDA}
\end{equation}
\end{prop}

By the relations (\ref{CB}) and (\ref{CBDA}) we get
\begin{equation} 
D_{m}(u)A_{m}(u+1)= A_{m+1}(u+1) A_{m-1}(u) - B_{m}(u)C_{m}(u+1).
\label{DAAA}
\end{equation}

By Proposition \ref{pr1}, the coefficients $ A_m^{(s)} $ of the series
 $A_{1}(u), \dots A_{n}(u)$:
\begin{equation}
A_m(u) = \sum_{s\geq 0} u^{-s} A_m^{(s)} \qquad ( m=1,2,\dots,n)
\end{equation}
 generate the commutative sub-algebra in $Y(\gln)$. 
 This algebra is called Gelfand-Zetlin algebra and is denoted by $A(\gln )$.

The following lemma will be used in the next section:
\begin{lemma} Let $\quad \mbox{}^*$ : $Y(\gln) \rightarrow Y(\gln)$ be the algebra anti-involution such that 
\begin{gather}
{T_{a,b}^{(1)}}^* = T_{b,a}^{(1)}, \quad {T_{a,b}^{(2)}}^* = T_{b,a}^{(2)}, \quad \text{and} \quad  {A_n^{(t)}}^* =  A_n^{(t)} \qquad (t=0,1,2,\dots).\\ 
\intertext{Then}
{T_{a,b}^{(s)}}^* = T_{b,a}^{(s)} \qquad \text{for all $ s = 0,1,2,\dots \; $.}
\end{gather} \label{lemma:anti-involution}
\end{lemma}  
\begin{pf}
The lemma is proven by induction in the $s$. Suppose ${T_{a,b}^{(r)}}^* = T_{b,a}^{(r)}$ hold for all $r \leq s $. Then the relations of the Yangian (\ref{eq:Yangianrel}) and 
\begin{equation}
{T_{a,b}^{(1)}}^* = T_{b,a}^{(1)}, \quad {T_{a,b}^{(2)}}^* = T_{b,a}^{(2)} \label{eq:inductbaselemma1}
\end{equation}
entail 
\begin{equation}
{T_{a,b}^{(s+1)}}^* = T_{b,a}^{(s+1)} \quad (a\neq b) \qquad \text{and} \qquad 
\left( T_{a,a}^{(s+1)} - T_{b,b}^{(s+1)}\right)^* =  T_{a,a}^{(s+1)} - T_{b,b}^{(s+1)}.
\end{equation}
And the condition on the quantum determinant:
\begin{equation}
{A_n^{(t)}}^* =  A_n^{(t)} \qquad (t=0,1,2,\dots \;)
\end{equation}
gives 
\begin{equation}
\left(T_{1,1}^{(s+1)} + T_{2,2}^{(s+1)}+ \cdots + T_{n,n}^{(s+1)}\right)^* = T_{1,1}^{(s+1)} + T_{2,2}^{(s+1)}+ \cdots + T_{n,n}^{(s+1)}.
\end{equation}
This completes the proof of the induction step. Taking the conditions (\ref{eq:inductbaselemma1}) as the induction base we obtain the statement of the lemma.
\end{pf}

\subsection{Yangian Gelfand-Zetlin bases}

Let $V$ be an irreducible finite dimensional $\gln$-module and $E_{a,b}$ be the generators of $\gln$.
 Denote by $v_{\l }$ the  highest weight vector in $V$ :
\begin{equation}
E_{a,a}\cdot v_{\l }=\l_a v_{\l } \qquad E_{a,b}\cdot v_{\l } =0,\quad a<b.
\end{equation}
Then each difference $\l_a-\l_{a+1}$ is a non-negative integer.
 We assume that each $\l_a$ is also an integer.
Denote by ${\cal T}_{\l } $ the set of all arrays $\Lambda$
 with integral entries of the form
\begin{eqnarray}
& \l _{n,1} \; \; \l _{n,2} \; \cdots \cdots \cdots \cdots \cdots \; \; 
 \l _{n,n} &  \\
& \l _{n-1,1} \; \;   \cdots \cdots \; \; \l _{n-1,n-1} &
\nonumber \\
& \ddots \; \; \cdots \cdots \; \; \; \; \; \; \;  & \nonumber \\
& \l _{2,1} \; \; \l _{2,2} & \nonumber \\
& \l _{1,1} & \nonumber
\end{eqnarray}
where $\l _{n,i}=\l _i$ and $\l _i \geq \l _{m,i}$ for all $i$ and $m$.
The array $\Lambda $ is called a  Gelfand-Zetlin scheme if
\begin{equation}
\l_{m,i}\geq\l_{m-1,i}\geq\l_{m,i+1}
\end{equation}
for all possible $m$ and $i$.
 Denote by ${\cal S}_{\l }$ the subset in ${\cal T}_{\l }$
consisting of the Gelfand-Zetlin schemes.

 There is a canonical decomposition of the space $V$ into the direct sum of
one-dimensional subspaces associated with the chain of sub-algebras
\begin{equation}
\gl_1 \subset \gl_2 \subset \ldots \subset \gln.
\end{equation}
These subspaces are parameterized by the elements $\Lambda \in {\cal S}_{\l }$.
 The subspace
$V_\Lambda \subset V$ corresponding to $\Lambda \in {\cal S}_{\l }$ 
is contained in an irreducible
$\gl_m$-submodule of the highest weight $(\l_{m,1},\l_{m,2},\ldots,\l_{m,m})$
for each $m=n-1,n-2,\ldots,1$. These conditions define $V_\Lambda$ uniquely.
\cite{gz}
 
\vspace{.2in}
Let us recall some facts about  representations of the Yangian $Y(\gln )$.

If we set $u'=u+h, v'=v+h \; \; (h \in \cplx )$, the relations (\ref{RTT})
 are also satisfied for $(u',v')$.
Thus the map
\begin{equation}
T_{a,b}(u) \mapsto T_{a,b}(u+h) \label{h}
\end{equation}
defines an automorphism of the algebra $Y(\gln )$.
So if there is a representation $V$ of $Y(\gln )$, we can construct another
representation of $Y(\gln )$ by the pullback through this automorphism.

We can regard the representation of the Lie algebra $\gln $
 as the representation of  $Y(\gln )$. This transpires  due to the existence of the  homomorphism $\pi_n$ from $Y(\gln )$ to $U(\gln )$:
 the universal enveloping algebra of $\gln $:
\begin{equation}
\pi_{n} \; : \; T_{a,b}(u) \mapsto \delta_{a,b} + E_{b,a}u^{-1}.
\end{equation}
 
 Let $V_{\l }$ be the irreducible $\gln $--module whose highest weight is 
$\l = (\l_{1} , \l_{2} , \dots ,\l_{n})$.
 We denote by $V_{\l}(h)$ the $Y(\gln )$--module obtained from $V_{\l }$
 by the pullback through this homomorphism and the automorphism (\ref{h}).

\vspace{.2in} 
The Yangian $Y(\gln )$ has the coproduct
 $\Delta :Y(\gln ) \rightarrow Y(\gln ) \otimes Y(\gln )$.
 It is given as follows.
\begin{equation}
\Delta (T_{a,b}(u)) = \sum_{c=1}^{n} T_{a,c}(u) \otimes T_{c,b}(u) .   \label{eq:coproduct}
\end{equation}
So if there are representations $V_{i} \; (i=1, \dots ,M)$
 of the Yangian $Y(\gln )$,
we can construct the representation $V_{1} \otimes V_{2} \otimes \cdots 
\otimes V_{M}$ of $Y(\gln )$:
\begin{equation}
T_{a,b}(u) \cdot (v_{1} \otimes v_{2} \otimes \cdots \otimes v_{M}) =
\Delta ^{(n-1)}\circ \cdots \circ \Delta ^{(2)} 
(T_{a,b}(u)) (v_{1} \otimes v_{2} \otimes \dots \otimes v_{M} ) 
\end{equation}
\begin{equation}
= \sum_{k_{1} \dots k_{M-1}} T_{a,k_{1}}(u) v_{1} \otimes 
T_{k_{1},k_{2}}(u) v_{2} \otimes \cdots \otimes T_{k_{M-1},b}(u) v_{M}.
\nonumber
\end{equation}

\mbox{} From now on
 we consider the following representation of the Yangian $Y(\gln )$: 
\begin{equation}
W= V_{\lambda ^{(1)}}(h^{(1)}) \otimes  V_{\lambda ^{(2)}}(h^{(2)}) \otimes 
\cdots \otimes  V_{\lambda ^{(M)}}(h^{(M)})
\end{equation}
where we assume that $h^{(r)}-h^{(s)} \not\in \zint$ for all $r \neq s$.

Let us  set $\rho _{0}(u)=1$ and for $m=1 ,\dots ,n$ let us  define
\begin{equation}
\rho _{m}(u) = \prod _{s=1}^{M} \prod _{i=1}^{m} (u-i+1+h^{(s)}),
\end{equation}
and 
\begin{eqnarray}
& a_m(u)=\rho _m(u)A_m(u) \; & m=0, \cdots ,n \; ,\\
& b_m(u)=\rho _m(u)B_m(u) \; & m=1, \cdots ,n-1 \; , \label{bm} \\
& c_m(u)=\rho _m(u)C_m(u) \; & m=1, \cdots ,n-1 \; ,\\
& d_m(u)=\rho _m(u)D_m(u) \; & m=1, \cdots ,n-1 \; .
\end{eqnarray}
Then $a_m(u)$, $b_m(u)$, $c_m(u)$ and $d_m(u)$
are polynomials in $u$, and due to the proposition \ref{pr2}
 and (\ref{DAAA}), they satisfy
\begin{eqnarray}
 & [a_m(u),b_l(v)]=0 \; \; \mbox{if} \; \; \; \; l\neq m, & \\
 & [c_m(u),b_l(v)]=0  \; \; \mbox{if}  \; \; \; \;l\neq m, & \\
 & [b_m(u),b_l(v)]=0  \; \; \mbox{if}  \; \; \; \; |l-m|\neq1, & \label{bb} \\
 & (u-v) \cdot [a_m(u),b_m(v)]=b_m(u) a_m(v)-b_m(v) a_m(u), & \\
 & (u-v) \cdot [c_m(u),b_m(v)]=d_m(u) a_m(v)-d_m(v) a_m(u), & \label{cb} \\
 & d_m(u) a_m(u+1) = a_{m+1}(u+1)a_{m-1}(u) - b_m(u)c_m(u+1). & \label{daaa}
\end{eqnarray} 

Let us fix a set of Gelfand-Zetlin schemes 
\begin{equation}
\Lambda ^{(s)} = ( \lambda _{m,i} | 1 \leq i \leq m \leq n)
 \in {\cal T}_{\l ^{(s)}} \; \; \; (s=1, \dots ,M),
\end{equation}
and define the following polynomials for $m=0, \cdots ,n$. 
\begin{equation}
\varpi _{m, \Lambda ^{(1)}, \dots ,\Lambda ^{(M)}}(u)=
\prod _{s=1}^{M} \prod _{i=1}^{m}(u+\l ^{(s)}_{m,i}-i+1+h^{(s)}).
\end{equation}
Note that all the zeroes of
the $m$--th polynomial
\begin{equation}
\nu_{m,i}^{(s)}=i-\l _{m,i}^{(s)}-1-h^{(s)},
\end{equation}
are pairwise distinct due to our assumption on the parameters
$h^{(1)}, \dots , h^{(M)}$.

For the pairs $(m,m') \; (  1 \leq m' \leq m \leq n)$, we introduce the
ordering,
\begin{equation}
(m,m') \prec (l,l') \; \; \Leftrightarrow \; \; 
m'<l' \mbox{ or } ( \; m'=l' \mbox{ and } m>l ).
\end{equation}
Let $v_{h.w.v} \in W$ be the vector, which is the tensor product of
the highest weight vectors $v^{(s)}_{h.w.v}$ of the Lie algebra
 $\gln $ $(s=1, \cdots ,M)$.
Then consider the following vector in $W$
\begin{equation}
v_{ \Lambda ^{(1)}, \dots ,\Lambda ^{(M)}}  =
\prod_{(m,m')}^\rightarrow\
\left( \prod_{(s,t) \atop{1 \leq t \leq \l ^{(s)}_{n,m'}- \l ^{(s)}_{m,m'}}}
b_m(\nu_{m,m'}^{(s)}-t)
\right) \cdot v_{h.w.v},
\end{equation}
Here for each fixed $m$ the elements
$b_m(\nu_{m,m'}^{(s)}-t) \in End(W)$ commute
 because of the relation (\ref{bb}).

Then the following propositions are satisfied. (See \cite{nt2}) 
\begin{prop} {\em \cite{nt2}} \label{am}
For every $m=1, \cdots , n$ we have the equality
\begin{equation}
a_m(u) \cdot v_{ \Lambda ^{(1)}, \dots ,\Lambda ^{(M)}}
 =\varpi_{m,\Lambda ^{(1)} , \dots ,\Lambda^{(M)}}(u)
 \cdot v_{ \Lambda ^{(1)}, \dots ,\Lambda ^{(M)}}.
\end{equation}
\end{prop}
\begin{prop}{\em \cite{nt2}}  \label{ze}
If $\Lambda ^{(r)} \notin {\cal S}_{\l^{(r)}}$ for some $r\in\{1, \dots , M\}$,
 then $v_{ \Lambda ^{(1)}, \dots ,\Lambda ^{(M)}}=0$.
\end{prop}
\begin{prop}{\em \cite{nt2}}  \label{nze}
If $\Lambda ^{(r)} \in {\cal S}_{\l^{(r)}}$ for every $r\in\{1, \dots , M\}$,
 then $v_{ \Lambda ^{(1)}, \dots ,\Lambda ^{(M)}} \neq 0$.
\end{prop}
\begin{prop}{\em \cite{nt2}}  \label{irr}
$Y(\gln )$-module $W$ is irreducible
if $h^{(r)}-h^{(s)} \notin \zint $ for all $r \neq s$.
\end{prop}
By the propositions \ref{am}, \ref{nze} and the fact that if 
$(\Lambda ^{(1)} , \dots ,\Lambda^{(M)}) \neq 
(\tilde{\Lambda }^{(1)} , \dots ,\tilde{\Lambda}^{(M)})$ 
 $(\forall r , \; \Lambda ^{(r)} , \;  \tilde{\Lambda}^{(r)}
 \in {\cal S}_{\l^{(r)}})$
 then $\exists m $ s.t. 
$\varpi_{m,\Lambda ^{(1)} , \dots ,\Lambda^{(M)}}(u) \neq  
\varpi_{m,\tilde{\Lambda }^{(1)} , \dots ,\tilde{\Lambda}^{(M)}}(u) $,
 one can show 
\begin{prop} \label{base}
$v_{ \Lambda ^{(1)}, \dots ,\Lambda ^{(M)}} \; \; 
(\Lambda ^{(r)} \in {\cal S}_{\l^{(r)}}$ for every $r\in\{1, \dots , M\}$) 
form a  base of $W$.
\end{prop}

\label{ss:GZ}

\section{Yangian in the Spin Calogero-Sutherland Model} \label{sec:yangian}

In this section we recall the definition of the  Yangian action in  the SCSM  \cite{BGHP} and establish some properties of this action -- in particular the self-adjointness of the operators giving the action of the Gelfand-Zetlin algebra (proposition \ref{p:sa}).  

Following \cite{BGHP} for $\kappa =\pm $ define the Monodromy operator $\hat{T}^{(\kappa)}_0(u)$ $\in End(\cplxn)\otimes End(\h)[[u^{-1}]]$ by  
\begin{equation}
\hat{T}^{(\kappa)}_0(u) = \sum_{a,b =1}^n E_{a,b}\otimes \hat{T}^{(\kappa)}_{a,b}(u) := \left(1 +\frac{P_{0,1}}{u -\kappa d_1}\right)\left(1 + \frac{P_{0,2}}{u -\kappa d_2}\right)\ldots\left(1 + \frac{P_{0,N}}{u -\kappa d_N}\right) \label{eq:That}
\end{equation}
the $P_{0,i}$ in this definition is the permutation operator of the 0-th and $i$-th tensor factors $\cplxn$ in the tensor product 
\begin{equation}
\underset{0}{\cplxn}\otimes \cz \otimes \underset{1}{\cplxn}\otimes \underset{2}{\cplxn}\otimes \cdots \otimes \underset{N}{\cplxn}  = \underset{0}{\cplxn}\otimes \h.
\end{equation}
The $E_{a,b} \in End(\cplxn)$ is the standard matrix unit in the basis $\{ v_{\ep}\}$ introduced before the definition (\ref{eq:sps}). 
The operators $ \hat{T}_{a,b}^{(\kappa),(s)} \in End(\h)$ obtained by expanding the Monodromy matrix  $\hat{T}^{(\kappa)}_{a,b}(u)$:    
\begin{equation}
\hat{T}^{(\kappa)}_{a,b}(u) = \delta_{a,b}1  +  \sum_{s\geq 1} u^{-s}  \hat{T}_{a,b}^{(\kappa),(s)}
\end{equation}
satisfy the defining relations (\ref{eq:Yangianrel}) of the $Y(\gln)$. By using the relations of the degenerate affine Hecke algebra (\ref{eq:Hecke1}-\ref{eq:Hecke3}) one can show \cite{BGHP} that the $\hat{T}_{a,b}^{(\kappa),(s)}$ leave the subspace $\hs{\kappa}$ invariant. We will set  
\begin{equation}
T^{(\kappa)}_{a,b}(u) := \left.\hat{T}^{(\kappa)}_{a,b}(u)\right|_{\hs{\kappa}} \quad \in End(\hs{\kappa})[[u^{-1}]] \qquad (a,b =1,2,\dots,n).  \label{eq:T}
\end{equation}

Denote the generating series  which give the action of the Gelfand-Zetlin algebra in the Yangian representation defined by the Monodromy matrix (\ref{eq:T}) by $ A_1^{(\kappa)}(u), A_2^{(\kappa)}(u),\dots , A_n^{(\kappa)}(u) $.
The $A^{(\kappa)}_n(u)$ is just the quantum determinant of the $T^{(\kappa)}_{a,b}(u)$. Hence
\begin{equation}
[A^{(\kappa)}_n(u), T^{(\kappa)}_{a,b}(v)] = 0 \qquad (a,b =1,2,\dots,n).
\end{equation}
The explicit expression for the quantum determinant  \cite{BGHP}:
\begin{equation}
A^{(\kappa)}_n(u) = \prod_{i=1}^N \frac{u + 1 -\kappa d_i}{u -\kappa d_i} \label{eq:qdet}
\end{equation} 
shows that the SCSM Hamiltonian (\ref{eq:H})is an element in the center of the Yangian action and hence is an element in the Gelfand-Zetlin algebra.

Denote by $O^{\dagger}$ the adjoint of an operator $O \in End(\hs{\kappa})$ with respect to the scalar product  ${\sprod{\:\cdot\:}{\:\cdot\:}}_{(\kappa)}$ defined in sec. \ref{sec:definition}. For $O(u)=\sum_{s\geq 0}u^{-s}O^{(s)}$ $\in$ $End(\hs{\kappa})[[u^{-1}]]$ we will write $O(u)^{\dagger} := \sum_{s\geq 0}u^{-s}O^{(s) {\dagger}}$.

\begin{prop}  \begin{equation}  {T_{a,b}^{(\kappa)}(u)}^{\dagger} = T_{b,a}^{(\kappa)}(u)  \qquad (\kappa = -,+). \end{equation} 
\end{prop}
\begin{pf} By lemma \ref{lemma:anti-involution} to prove the proposition  it is sufficient to show that  
\begin{gather}
{T_{a,b}^{(\kappa),(1)}}^{\dagger} = T_{b,a}^{(\kappa),(1)}, \qquad  {T_{a,b}^{(\kappa),(2)}}^{\dagger} = T_{b,a}^{(\kappa),(2)} \label{eq:p91}\\ 
\intertext{and}
{A_n^{(\kappa)}(u)}^{\dagger} = A_n^{(\kappa)}(u).\label{eq:p92}
\end{gather}
Using the definition (\ref{eq:T}) and the same notation regarding the subscript $ _0$ as in (\ref{eq:That}) we can write
\begin{gather*}
 T_{0}^{(\kappa),(1)} = \sum_{i=1}^N P_{0,i},  \\
 T_{0}^{(\kappa),(2)} = \left.\left(\sum_{i=1}^N \kappa d_i P_{0,i} + \sum_{1\leq i < j \leq N} P_{0,i}P_{0,j} \right)\right|_{\hs{\kappa}} = \left.\left(\sum_{i=1}^N \kappa d_i P_{0,i} + \sum_{1\leq i < j \leq N} \kappa K_{i,j}P_{0,j} \right)\right|_{\hs{\kappa}}
\end{gather*}
The Dunkl operators $d_i$ and the permutation operators $K_{i,j}$ $(i,j=1,2,\dots,N)$ are self-adjoint with respect to the scalar product (\ref{eq:spc}). On the other hand for any $x, y \in \Ncplxn $ we have
\begin{equation}
{\sprod{\: P_{0,i}\, x\:}{\:y\:}}_s  = {\sprod{\: x\:}{\:P_{0,i}^{t_0}\,y\:}}_s
\end{equation}
where $\mbox{}^{t_0}$ stands for the matrix transposition in the auxiliary space $\cplxn$ (\ref{eq:That}). Using the definitions of the scalar products (\ref{eq:sp}) and ${\sprod{\: \cdot\:}{\:\cdot\:}}_{(\kappa)}$ we obtain  (\ref{eq:p91}).

The (\ref{eq:p92}) follows from the explicit expression for the quantum determinant (\ref{eq:qdet}) and the self-adjointness of the Dunkl operators with respect to the scalar product (\ref{eq:sp}).
\end{pf}

Using this Proposition we can now establish the main result of this section:
\begin{prop}
\begin{equation}
{A_{m}^{(\kappa)}(u)}^{\dagger }= A_m^{(\kappa)}(u), \; \; \; {B_{m}^{(\kappa)}(u)}^{\dagger }= C_m^{(\kappa)} (u)
  , \; \; \; {C_m^{(\kappa)}(u)}^{\dagger }= B_m^{(\kappa)} (u) \qquad (\kappa = -,+).
\end{equation}  \label{p:sa}
\end{prop}
\begin{pf}
Since in the following proof it is immaterial whether we are dealing with the fermionic or bosonic case, we will suppress the superscripts $(\kappa)$. 

In the paper \cite{MNO}, the proof of the following relations can be found:
\begin{equation}
T_{i_1,, j_1}(u) T_{i_2, j_2}(u-1) \cdots T_{i_m, j_m}(u-m+1) 
E_{i_1, j_1}^{(1)} E_{i_2, j_2}^{(2)} \cdots E_{i_m, j_m}^{(m)}
( H_{m} \otimes 1) \label{anti}
\end{equation}
\begin{equation}
= ( H_{m} \otimes 1)
 E_{i_1, j_1}^{(1)} E_{i_2,j_2}^{(2)} \cdots E_{i_m, j_m}^{(m)}
 T_{i_m, j_m}(u-m+1) \cdots T_{i_2, j_2}(u-1) T_{i_1, j_1}(u) 
\nonumber
\end{equation}
The relations (\ref{anti}) are satisfied in $End(\cplx ^{n}) ^{\otimes m}
\otimes Y(\gln )[[u^{-1}]]$, and $H_{m} \in End(\cplx ^{n}) ^{\otimes m}$ is the
antisymmetrization map. 
By comparing the coefficient of
 $ E_{i_1, j_1}^{(1)} E_{i_2, j_2}^{(2)} \cdots E_{i_m, j_m}^{(m)}$,
we get
\begin{equation}
\sum_{\sigma \in {\frak S}_{m}} (-1)^{l(\sigma )} 
T_{i_1, j_{\s (1)}}(u) T_{i_2, j_{\s (2)}}(u-1) \cdots T_{i_m, j_{\s (m)}}(u-m+1)
\end{equation}
\begin{equation} 
=\sum_{\sigma \in {\frak S}_{m}} (-1)^{l(\sigma )}
T_{i_{\s (m)}, j_m}(u-m+1)\cdots T_{i_{\s (2)}, j_2}(u-1) T_{i_{\s (1)}, j_1}(u). 
\nonumber
\end{equation}
Then if we take the adjoint, we have
\begin{equation}
( \sum_{\sigma \in {\frak S}_{m}} (-1)^{l(\sigma )} 
T_{i_1, j_{\s (1)}}(u) T_{i_2, j_{\s (2)}}(u-1) \cdots
 T_{i_m, j_{\s (m)}}(u-m+1) 
) ^{\dagger } 
\end{equation}
\begin{equation} 
=\sum_{\sigma \in {\frak S}_{m}} (-1)^{l(\sigma )} 
T_{j_1, i_{\s (1)}}(u) T_{j_2, i_{\s (2)}}(u-1) \cdots
 T_{j_m, i_{\s (m)} }(u-m+1) 
\nonumber
\end{equation}
If we put $(i_1 , \dots i_m) = (1, \dots , m) , \; \; 
 (j_1 , \dots j_m) = (1, \dots , m)$,
 we get $A_m(u)^{\dagger }=A_m(u)$,
and if we put $(i_1 , \dots i_m) = (1, \dots , m) , \; \; 
 (j_1 , \dots j_m) = (1, \dots , m-1, m+1)$ (resp.  $(i_1 , \dots i_m)
 = (1, \dots , m-1,m+1) , \; \;  (j_1 , \dots j_m) = (1, \dots ,m)$ ),
we get $B_m(u)^{\dagger }=C_m(u)$ (resp. $C_m(u)^{\dagger }=B_m(u)$ ).
\end{pf}

In sec. \ref{sec:dnorms} we will see that the  operator coefficients generated by the $A^{(\kappa)}_1(u),\dots,A^{(\kappa)}_n(u)$ are simultaneously diagonalizable in $\hs{\kappa}$, and that their joint spectrum is multiplicity free. Since the  $A^{(\kappa)}_1(u),\dots,A^{(\kappa)}_n(u)$ are self-adjoint this implies that their common eigenvectors are mutually orthogonal with respect to the scalar product ${\sprod{\:\cdot\:}{\:\cdot\:}}_{(\kappa)}$ . Our main problem in this paper is to describe these eigenvectors and to compute their norms.

\section{Decomposition of the space of states into irreducible Yangian submodules} \label{sec:decomposition}

In this section we construct the decomposition of the space of states of SCSM into irreducible submodules of the Yangian action. The procedure we follow is the one suggested in \cite{BGHP},  it is based on the diagonalization of the Dunkl operators. The eigenvectors of the Dunkl operators, known as non-symmetric Jack polynomials, are reviewed in the subsection \ref{ssec:nsjack}. In the subsection \ref{ssec:fdec} we describe the decomposition of the space of states $\hs{-}$ in the fermionic Model, the main result here is the theorem \ref{t:FDEC}. In the subsection  \ref{ssec:bdec} we give the decomposition in the bosonic case.   

\subsection{Non-symmetric Jack Polynomials} \label{ssec:nsjack}
In this subsection we consider the Dunkl operators (\ref{eq:Dunkl}) as acting  in $\cz$. For $\alpha > 0$ the Dunkl operators are simultaneously diagonalizable. Their common eigenvectors form a base in $\cz$ and are sometimes called non-symmetric Jack polynomials. Here we will review some of the properties of these polynomials.

First we describe the  labeling of the eigenvectors which will be convenient  in the proofs of the statements we are going to make later. Let $\MC_N := $ $\{ (\mlett_1,\mlett_2,\dots,\mlett_N) \in \zint^N \; | \; \mlett_1 \geq \mlett_2 \geq \dots \geq \mlett_N \}$ be the set of partitions which may have negative parts. There is a right action of the symmetric group $\Sgroup{N}$ in $\zint^N$. For $\s =$ \begin{small}$\left(\ba{cccc} 1 & 2 & \cdots & N \\ \s(1) & \s(2) & \cdots & \s(N) \ea \right)$ \end{small} $\in \Sgroup{N}$ and $(\nlett_1,\nlett_2,\dots,\nlett_N) \in \zint^N $ it is defined by     
\begin{equation}
\s.(\nlett_1,\nlett_2,\dots,\nlett_N) = (\nlett_{\s(1)},\nlett_{\s(2)},\dots,\nlett_{\s(N)}).
\end{equation}
For an $\mm \in \MC_N$  we define the subset $S^{\mm}$ in $\Sgroup{N}$ by 
\begin{gather}
\s \in S^{\mm} \quad \text{iff}\label{eq:Slambda} \\ \text{ for all $1\leq i \leq N $} \quad \s(i) = \#\{\;j\leq i \;| \; \mlett_{\s(j)} \geq  \mlett_{\s(i)}\} + \#\{\;j>i \;| \; \mlett_{\s(j)} >  \mlett_{\s(i)}\}.  \nonumber
\end{gather}
Let ${\frak S}^{\mm}_N$ $\subset$ $\Sgroup{N}$ be the subgroup leaving the $\mm$ invariant.  Then $S^{\mm}$ intersects each of the right cosets of  ${\frak S}^{\mm}_N$ in $\Sgroup{N}$ at precisely one element, and the correspondence between $S^{\mm}$  and the set of all distinct rearrangements of the $\mm$ given by    
\begin{equation}
 \s \in S^{\mm} \; \rightarrow \; \s.\mm = (\mlett_{\s(1)},\mlett_{\s(2)},\dots,\mlett_{\s(N)}) 
\end{equation}
is bijective.

Some of the properties of the set $S^{\mm}$  are summarized as follows:  
\begin{align}
 & \text{if $\s \in S^{\mm}\quad$    then $\;\s(i,i+1)\in S^{\mm}$ iff $\mlett_{\s(i)} \neq \mlett_{\s(i+1)}$.}  \label{eq:p1}  \\   
 & \text{if $\s \in S^{\mm}\;$ then $\;l(\s) \left(:= \sum_{i<j}\theta(\s(i) > \s(j))\right) = \sum_{i<j}\theta(\mlett_{\s(i)} < \mlett_{\s(j)}) $.} \label{eq:p2} \\  
 & \text{ $\forall$ $\s \in S^{\mm}$, $\s \neq {\id}$ $\;\exists\;$ $(i,i+1)$ such that $\mlett_{\s(i)} < \mlett_{\s(i+1)}$ and $l(\s(i,i+1)) = l(\s) - 1$. }  \label{eq:p3}
\end{align} 
Here in the definition of the length $l(\s)$ we used the convention $ \theta(x) = 1 $ if $x$ is true , $ \theta(x) = 0 $ if $x$ is false.

In the set $S^{\mm}$  introduce the total ordering by setting
\begin{gather}
\s \succ \s'  \label{eq:Sord}\\  \text{iff } \quad \text{the last non-zero element in $(\mlett_{\s(1)} - \mlett_{\s'(1)},\mlett_{\s(2)}-\mlett_{\s'(2)},\dots,\mlett_{\s(N)}-\mlett_{\s'(N)})$ is $< 0$.} \nonumber
\end{gather}
Notice that the identity in $\Sgroup{N}$ is the maximal element in $S^{\mm}$ in this ordering. 
Then in the set of pairs $(\mm,\s)$ $(\mm \in \MC_N, \; \s \in S^{\mm})$ the partial ordering is defined by 
\begin{equation}
(\mm,\s) > (\tilde{\mm},\tilde{\s}) \quad \text{iff} \quad \begin{cases} \mm > \tilde{\mm} & \text{or} \\
\mm = \tilde{\mm}, &  \s \succ \tilde{\s} \end{cases}
\end{equation}
where $\mm > \tilde{\mm}$ means that $\mm$ is greater than $\tilde{\mm}$ in the dominance (natural) ordering in $\MC_N$ \cite{Macdonaldbook}.

The eigenvectors $\Phi_{\s}^{\mm}(z) \in \cz$ of the Dunkl operators are labeled by the pairs $(\mm,\s)$ $(\mm \in \MC_N, \; \s \in S^{\mm})$ and satisfy the following properties: 
\begin{align}
& \Phi_{\s}^{\mm}(z) = z_1^{\mlett_{\s(1)}}z_2^{\mlett_{\s(2)}}\cdots z_N^{\mlett_{\s(N)}} + \sum_{(\tilde{\mm},\tilde{\s}) < ({\mm},{\s})} c_{({\mm},{\s});(\tilde{\mm},\tilde{\s})} z_1^{\tilde{\mlett}_{\tilde{\s}(1)}}z_2^{\tilde{\mlett}_{\tilde{\s}(2)}}\cdots z_N^{\tilde{\mlett}_{\tilde{\s}(N)}}; \label{eq:triangularity}\\  
& d_i \Phi_{\s}^{\mm}(z) = \xi_i^{\mm}(\s) \Phi_{\s}^{\mm}(z), \quad \text{where}\quad \xi_i^{\mm}(\s):= \alpha \mlett_{\s(i)} - \s(i) \qquad (i=1,2,\dots,N); \label{eq:d-eigenvectors}\\
& K_{i,i+1}\Phi_{\s}^{\mm}(z) = {\cal A}_i^{\mm}(\s)\Phi_{\s}^{\mm}(z) + {\cal B}_i^{\mm}(\s)\Phi_{\s(i,i+1)}^{\mm}(z),   \label{eq:KactionPhi1}\\ 
\intertext{where}
& {\cal A}_i^{\mm}(\s) = \frac{1}{ \xi_i^{\mm}(\s) -  \xi_{i+1}^{\mm}(\s)}, \quad  {\cal B}_i^{\mm}(\s) = \begin{cases} \frac{\left(\xi_i^{\mm}(\s) -  \xi_{i+1}^{\mm}(\s)\right)^2 - 1}{\left(\xi_i^{\mm}(\s) -  \xi_{i+1}^{\mm}(\s)\right)^2} & (\mlett_{\s(i)} >  \mlett_{\s(i+1)}) , \\ \qquad  0  &  (\mlett_{\s(i)} =  \mlett_{\s(i+1)}), \\
\qquad 1 & (\mlett_{\s(i)} <  \mlett_{\s(i+1)}). \end{cases} \label{eq:KactionPhi2}
\end{align}
Notice that for $\s \in S^{\mm}$ we have $\s(i+1) = \s(i)+1$ whenever $\mlett_{\s(i)} =  \mlett_{\s(i+1)}$, and hence in this case the  (\ref{eq:KactionPhi1}, \ref{eq:KactionPhi2}) give \begin{equation}
K_{i,i+1}\Phi_{\s}^{\mm}(z) = \Phi_{\s}^{\mm}(z) \qquad (\mlett_{\s(i)} =  \mlett_{\s(i+1)}). \label{eq:Kphi=phi}
\end{equation}

For $\alpha > 0$, the set of $N$ eigenvalues $ (\xi_1^{\mm}(\s),\xi_2^{\mm}(\s), \dots,  \xi_N^{\mm}(\s) )$ determines the pair  $(\mm,\s)$ uniquely. Since the Dunkl operators are self-adjoint with respect to the scalar product ${\sprod{\;\cdot\;}{\;\cdot\;}}_c$ (\ref{eq:spc}) this implies that the eigenvectors $\Phi_{\s}^{\mm}(z) $  are mutually orthogonal:
\begin{equation}
{\sprod{\Phi_{\s}^{\mm}(z)}{\Phi_{\kappa}^{\nn}(z)}}_c  = \delta_{\mm,\nn}\delta_{\s,\kappa} \| \Phi_{\s}^{\mm}(z)\|^2_c. 
\end{equation}
Their norms  $\| \Phi_{\s}^{\mm}(z)\|^2_c$ have been computed in \cite{Opdam} and for the $q$-deformed situation in \cite{Macdonald1,Cherednik2}. The product formulas for the norms  $\| \Phi_{{\id}}^{\mm}(z)\|^2_c$ will be used in sec. \ref{sec:hwvnorms} to derive  product formulas for the norms of the Yangian highest-weight vectors.

\subsection{Irreducible decomposition of the space of states with respect to the Yangian action. Fermionic case}\label{ssec:fdec}
In this subsection we describe the decomposition of the space of states in the fermionic SCSM: $\hs{-}$ into irreducible subrepresentations with respect to the $Y(\gln)$-action (\ref{eq:T}) $(\kappa = -)$. 

Let  $ E^{\mm} := \oplus_{\s\in S^{\mm}}\cplx\Phi_{\s}^{\mm}(z) $ $(\mm \in \MC_N)$. And let  
\begin{equation}
F^{\mm} := ( E^{\mm}\otimes (\Ncplxn))\cap \hs{-}.  \label{eq:Flambda}
\end{equation}
The (\ref{eq:d-eigenvectors}) implies that the space $ F^{\mm}$ is invariant with respect to the Yangian action defined by (\ref{eq:T}) with $\kappa = -$. And since the polynomials   $\Phi_{\s}^{\mm}(z)$ $(\mm \in \MC_N, \; \s \in S^{\mm})$  form a base in $\cz$ we have 
\begin{equation}
\hs{-} = \bigoplus_{\mm \in \MC_N} F^{\mm}.
\end{equation}
The expression (\ref{eq:qdet}) ($\kappa = -$) implies that, unless $F^{\mm} = \emptyset $, $F^{\mm}$  is an eigenspace of the quantum determinant with the eigenvalue
\begin{equation}
\prod_{i=1}^N\frac{u + 1 + \xi_i^{\mm}({\id})}{u + \xi_i^{\mm}({\id})}
\end{equation}
and hence is an eigenspace of the Hamiltonian $H^{(-)}$ (\ref{eq:H}).

To describe each of the components $ F^{\mm}$ explicitly we need to make several definitions.

Let $W^{\mm}_{(-)} \subset \Ncplxn $  $(\mm \in \MC_N)$ be defined by 
\begin{equation}
W^{\mm}_{(-)} := \bigcap_{1\leq i \leq N \; \text{s.t.} \; \mlett_i = \mlett_{i+1}} Ker( P_{i,i+1}+ 1). \label{eq:W}
\end{equation}
Note that $\dim W^{\mm}_{(-)}=0$ unless $\mm \in \MC_N^{(n)}$ where 
\begin{equation}
\MC_N^{(n)}:= \{\; \mm \in \MC_N \;|\; \#\{\;\mlett_k\; |\;\mlett_k = i\;\} \leq n\quad (i\in \zint)\}. \label{eq:nstrict}\end{equation}

For $p \in \{1,2,\dots,n\}$ let $\la$ be the highest weight of the fundamental $\gln$-module:
\begin{equation}
\la = (\underbrace{1,1,\dots,1}_{p},\underbrace{0,0,\dots,0}_{n-p})\qquad ( 1 \leq p \leq n ). 
\end{equation}
For a highest weight of this  form and $h \in \cplx$ denote the corresponding $Y(\gln)$-module $V_{\la}(h)$ (see subsection \ref{ss:GZ}) by $V_p(h)$. As a linear space the  $V_p(h)$ is realized as the totally antisymmetrized tensor product of $\cplxn$:
\begin{equation}
V_p(h) =  \cap_{i=1}^{p-1} Ker(P_{i,i+1}+1) \quad \subset \quad \otimes^p\cplxn \qquad  (1\leq p \leq n)
\end{equation}
with normalization chosen so that the $\gln$ highest weight vector in  $V_p(h)$ is 
\begin{equation}
{\omega}_p := \sum_{\s \in \Sgroup{p}} (-1)^{l(\s)} v_{\s(1)}\otimes v_{\s(2)} \otimes \cdots \otimes v_{\s(p)} . \label{eq:baromegap}
\end{equation}

For an $\mm \in \MC_N^{(n)}$ let $M$ be the number of distinct elements in the sequence 
\begin{equation}
\mm = (\mlett_1,\mlett_2,\dots,\mlett_N). \nonumber
\end{equation}
And let $ p_s $ $( 1\leq p_s \leq n, \quad s=1,2,\dots,M)$ be the multiplicities of the elements in the $\mm$:   
\begin{multline}
\mlett_1=\mlett_2=\cdots =\mlett_{p_1} > \mlett_{1+p_1}= \mlett_{2+p_1}=\cdots =\mlett_{p_2+p_1}> \quad \cdots \\ \cdots > \mlett_{1+ p_{M-1}+\cdots +p_2+p_1}=  \mlett_{2+ p_{M-1}+\cdots +p_2+p_1}=\cdots =\mlett_{p_{M}+\cdots +p_2+p_1 \equiv N}.  \label{eq:p_s}
\end{multline}
With $\xi^{\mm}_i := \xi^{\mm}_i({\id})$ (\ref{eq:d-eigenvectors}) set 
\begin{equation}
h_{\mm}^{(s)} := \xi^{\mm}_{1+p_1+p_2+\cdots+p_{s-1}} \qquad (p_0 :=0,\quad s=1,2,\dots,M ). \label{eq:h^s}
\end{equation}
Then for the linear space  $W^{\mm}_{(-)}$ (\ref{eq:W}) we have 
\begin{equation}
W^{\mm}_{(-)} =  
\begin{cases} {V}_{p_1}(h_{\mm}^{(1)})\otimes{V}_{p_2}(h_{\mm}^{(2)})\otimes\ \cdots \otimes {V}_{p_M}(h_{\mm}^{(M)}) \; \subset \; \otimes^N\cplxn  &  \text{ when $ \mm \in \MC_N^{(n)} $,} \\  
                     \qquad \emptyset \qquad  &  \text{ when $ \mm \not\in \MC_N^{(n)} $.}  \label{eq:W=V}
\end{cases}
\end{equation}
 When $ \mm \in \MC_N^{(n)} $ the $W^{\mm}_{(-)}$  is the Yangian module with the Yangian action defined by the coproduct (\ref{eq:coproduct}).

For any $\s \in S^{\mm}$ (\ref{eq:Slambda}) define $\check{\real}^{(-)}(\s)$ $\in$ $End(\Ncplxn)$ by the following recursion relation
\begin{gather}
\check{\real}^{(-)}({\id}) := 1, \\
\check{\real}^{(-)}(\s(i,i+1)) := -\check{R}_{i,i+1}\left( \xi_i^{\mm}(\s) - \xi^{\mm}_{i+1}(\s) \right)\check{\real}^{(-)}(\s) \qquad ( \;\mlett_{\s(i)} >\mlett_{\s(i+1)}\;)  \label{eq:Rrecursion}
\end{gather}
where the $R$-matrix is given by 
\begin{equation}
\check{R}_{i,i+1}(u) := u^{-1} + P_{i,i+1} \label{eq:Rcheck}
\end{equation}
Due to the property (\ref{eq:p3}) of the set $ S^{\mm}$ this recursion relation is sufficient to define $\check{\real}^{(-)}(\s)$  for all  $\s \in S^{\mm}$. The definition of the $\check{\real}^{(-)}(\s)$ is unambiguous by virtue of the Yang-Baxter equation satisfied by the  $R$-matrix (\ref{eq:Rcheck}).

For $\mm \in \MC_N$ define the map $U^{\mm}_{(-)}:$ $ \Ncplxn \rightarrow \h$ by setting for $v \in \Ncplxn$
\begin{equation}
U^{\mm}_{(-)}v := \sum_{\s \in S^{\mm}}\Phi_{\s}^{\mm}(z)\otimes \check{\real}^{(-)}(\s)v. \label{eq:U}
\end{equation}

\begin{thm}
For any $\mm \in \MC_N$ we have
\begin{equation}U^{\mm}_{(-)}: \: W^{\mm}_{(-)} \mapsto F^{\mm}. \end{equation}
 And the $U^{\mm}_{(-)}$ is an isomorphism of the $Y(\gln)$-modules $W^{\mm}_{(-)}$ and $F^{\mm}$. \label{t:FDEC}
\end{thm}
The proof of this theorem is given in the Appendix A.

This theorem  will allow us to use the results of \cite{nt2} described in sec. \ref{sec:background} in order to construct in $F^{\mm}$ the eigenbasis of the algebra $A(\gln)$ generated by the coefficients of the series $A^{(-)}_1(u),A^{(-)}_2(u),\dots,A^{(-)}_n(u)$. For now let us notice that from this theorem it follows that the Yangian highest weight vector $\Omega^{(-)}_{\mm}$ in $F^{\mm}$ is given by 
\begin{equation}
\Omega^{(-)}_{\mm} = U^{\mm}_{(-)}\omega_{\mm} = \sum_{\s \in S^{\mm}}\Phi_{\s}^{\mm}(z)\otimes \check{\real}^{(-)}(\s)\omega_{\mm} \label{eq:hwvector}
\end{equation}
where the $\omega_{\mm}$ is the highest weight vector in $W^{\mm}_{(-)}$:
\begin{equation}
\omega_{\mm} := {\o}_{p_1}\otimes{\o}_{p_2}\otimes \cdots \otimes{\o}_{p_M}. \label{eq:ol} \end{equation}

\mbox{} From the Corollary 3.9 in \cite{nt2} it follows that the modules $F^{\mm}$ are irreducible if $\alpha \not\in {\Bbb Q}$ since in this case in (\ref{eq:W=V}) we have $h^{(s)}_{\mm} -  h^{(r)}_{\mm} \not\in {\Bbb Z}$ when $s \neq r$. Using results of \cite{AK} we can verify, that the $F^{\mm}$ are irreducible under the weaker condition: $\alpha \in \real\setminus {\Bbb Q}_{\leq 0} $. The key statements of \cite{AK} which are used to come to this conclusion  are\\  
$ \bullet \; V_{p_1}(h^{(1)})\otimes V_{p_2}(h^{(2)}) $ is irreducible iff the $Y(\gln)$-intertwiner $R_{12} : V_{p_1}(h^{(1)})\otimes V_{p_2}(h^{(2)}) \rightarrow V_{p_2}(h^{(2)})\otimes V_{p_1}(h^{(1)})$ and the inverse intertwiner $R_{21}$ have no poles. \\
$ \bullet \; V_{p_1}(h^{(1)})\otimes V_{p_2}(h^{(2)})\otimes \cdots \otimes V_{p_M}(h^{(M)}) $ is irreducible iff $  V_{p_r}(h^{(r)})\otimes V_{p_s}(h^{(s)})$ is irreducible for all $ 1\leq r < s \leq M$.

\subsection{Irreducible decomposition of the space of states with respect to the Yangian action. Bosonic case} \label{ssec:bdec}
The decomposition of the space of states of the bosonic SCSM: $\hs{+}$ into irreducible sub-representations with respect to the $Y(\gln)$-action (\ref{eq:T}) $(\kappa = +)$ is carried out along the same lines as the one for the fermionic case.  

Let for $\mm \in \MC_N$ the $ E^{\mm} $ be defined as in the previous subsection. And let 
\begin{equation}
B^{\mm} := ( E^{\mm}\otimes (\Ncplxn))\cap \hs{+}.  \label{eq:Blambda}
\end{equation}
The (\ref{eq:d-eigenvectors}) implies that the space $ B^{\mm}$ is invariant with respect to the Yangian action defined by (\ref{eq:T}) with $\kappa = +$. And since the polynomials   $\Phi_{\s}^{\mm}(z)$ $(\mm \in \MC_N, \; \s \in S^{\mm})$  form a base in $\cz$ we have 
\begin{equation}
\hs{+} = \bigoplus_{\mm \in \MC_N} B^{\mm}.
\end{equation}
To describe each of the components $ B^{\mm}$ explicitly we make several definitions analogous to those made in the previous subsection.

Let $W^{\mm}_{(+)} \subset \Ncplxn $  $(\mm \in \MC_N)$ be defined by 
\begin{equation}
W^{\mm}_{(+)} := \bigcap_{1\leq i \leq N \; \text{s.t.} \; \mlett_i = \mlett_{i+1}} Ker( P_{i,i+1} - 1). \label{eq:W+}
\end{equation}

For $p =1,2,\dots $ let $\la$ be the following  $\gln$ highest weight:
\begin{equation}
\la = (p,\underbrace{0,0,\dots,0}_{n-1}). 
\end{equation}
For a highest weight of this  form and $h \in \cplx$ denote the corresponding $Y(\gln)$-module $V_{\la}(h)$ (see subsection \ref{ss:GZ}) by $V^p(h)$. As a linear space the  $V^p(h)$ is realized as the totally symmetrized tensor product of $\cplxn$:
\begin{equation}
V^p(h) =  \cap_{i=1}^{p-1} Ker(P_{i,i+1}-1) \quad \subset \quad \otimes^p\cplxn \qquad  (p=1,2,\dots).
\end{equation}
We choose normalization so that the highest weight vector in  $V_p(h)$ is equal to $ v_1^{\otimes p} $

As in the fermionic case,  for an $\mm \in \MC_N$ let $M$ be the number of distinct elements in the sequence $\mm = (\mlett_1,\mlett_2,\dots,\mlett_N)$. And let $ p_s $ $(   s=1,2,\dots,M)$ be the multiplicities of the elements in the $\mm$:   
\begin{multline}
\mlett_1=\mlett_2=\cdots =\mlett_{p_1} > \mlett_{1+p_1}= \mlett_{2+p_1}=\cdots =\mlett_{p_2+p_1}> \quad \cdots \\ \cdots > \mlett_{1+ p_{M-1}+\cdots +p_2+p_1}=  \mlett_{2+ p_{M-1}+\cdots +p_2+p_1}=\cdots =\mlett_{p_{M}+\cdots +p_2+p_1 \equiv N} .  \label{eq:p_s+}
\end{multline}
With $\xi^{\mm}_i := \xi^{\mm}_i({\id})$ (\ref{eq:d-eigenvectors}) set 
\begin{equation}
h_{\mm}^{(s)} := -\xi^{\mm}_{1+p_1+p_2+\cdots+p_{s-1}} \qquad (p_0 :=0,\quad s=1,2,\dots,M ). \label{eq:h^s+}
\end{equation}
Then for the linear space  $W^{\mm}_{(+)}$ (\ref{eq:W+}) we have 
\begin{equation}
W^{\mm}_{(+)} =  
 {V}^{p_1}(h_{\mm}^{(1)})\otimes{V}^{p_2}(h_{\mm}^{(2)})\otimes\ \cdots \otimes {V}^{p_M}(h_{\mm}^{(M)}) \; \subset \; \otimes^N\cplxn  \qquad ( \mm \in \MC_N^{(n)}).
\end{equation}
 The $W^{\mm}_{(+)}$  is the Yangian module with the Yangian action defined by the coproduct (\ref{eq:coproduct}).

For any $\s \in S^{\mm}$ (\ref{eq:Slambda}) define $\check{\real}^{(+)}(\s)$ $\in$ $End(\Ncplxn)$ by the following recursion relation
\begin{gather}
\check{\real}^{(+)}({\id}) := 1, \\
\check{\real}^{(+)}(\s(i,i+1)) := \check{R}_{i,i+1}\left( -\xi_i^{\mm}(\s) + \xi^{\mm}_{i+1}(\s) \right)\check{\real}^{(+)}(\s) \qquad ( \;\mlett_{\s(i)} >\mlett_{\s(i+1)}\;)  \label{eq:Rrecursion+}
\end{gather}
where the $R$-matrix $\check{R}_{i,i+1}(u)$ is given by (\ref{eq:Rcheck}).

As in the fermionic case, due to the property (\ref{eq:p3}) of the set $ S^{\mm}$ this recursion relation is sufficient to define $\check{\real}^{(+)}(\s)$  for all  $\s \in S^{\mm}$. The definition of the $\check{\real}^{(+)}(\s)$ is unambiguous by virtue of the Yang-Baxter equation satisfied by the  $R$-matrix (\ref{eq:Rcheck}).

For $\mm \in \MC_N$ define the map $U^{\mm}_{(+)}:$ $ \Ncplxn \rightarrow \h$ by setting for $v \in \Ncplxn$
\begin{equation}
U^{\mm}_{(+)}v := \sum_{\s \in S^{\mm}}\Phi_{\s}^{\mm}(z)\otimes \check{\real}^{(+)}(\s)v. \label{eq:U+}
\end{equation}

\begin{thm}
For any $\mm \in \MC_N$ we have
\begin{equation}U^{\mm}_{(+)}: \: W^{\mm}_{(+)} \mapsto B^{\mm}. \end{equation}
 And the $U^{\mm}_{(+)}$ is an isomorphism of the $Y(\gln)$-modules $W^{\mm}_{(+)}$ and $B^{\mm}$. \label{t:BDEC}
\end{thm}
We omit the proof of this theorem since it  is a straightforward modification of the proof of the theorem \ref{t:FDEC} given in Appendix A.
\mbox{}From this theorem it follows that the Yangian highest weight vector $\Omega^{(+)}_{\mm}$ in $B^{\mm}$ is given by 
\begin{equation}
\Omega^{(+)}_{\mm} = U^{\mm}_{(+)}v_1^{\otimes N} = \sum_{\s \in S^{\mm}}\Phi_{\s}^{\mm}(z)\otimes \check{\real}^{(+)}(\s)  v_1^{\otimes N} \label{eq:hwvector+}
\end{equation}
The theorem  \ref{t:BDEC} will allow us to use the results of \cite{nt2} summarized in sec. \ref{sec:background} in order to construct in $B^{\mm}$ the eigenbasis of the algebra $A(\gln)$ generated by the coefficients of the series $A^{(+)}_1(u),A^{(+)}_2(u),\dots,A^{(+)}_n(u)$. 

\section{Norms of the highest weight vectors in the irreducible Yangian submodules} \label{sec:hwvnorms}
\subsection{Fermionic case}
In this subsection we will compute the norms ${\sprod{\;\Omega^{(-)}_{\mm}\;}{\;\Omega^{(-)}_{\mm}\;}}_{(-)}$ of the highest weight vectors in each of the irreducible submodules $F^{\mm}$ $(\mm \in \MC_N^{(n)}$).

Let us fix an $\mm \in \MC_N^{(n)}$. In this subsection and later on we will use the notations (\ref{eq:p_s}, \ref{eq:h^s}). Let $\Phi^{\mm}(z) := \Phi_{{\id}}^{\mm}(z)$. Consider the vector 
\begin{equation}
\Ale^{(-)}_N \:\left( \Phi^{\mm}(z)\otimes\o_{\mm} \right)  \label{eq:APhio} 
\end{equation}
where $\Ale^{(-)}_N$ is the antisymmetrization  operator (\ref{eq:anti}). Due to (\ref{eq:KactionPhi1}) and the definition of the space $F^{\mm}$ (\ref{eq:Flambda}) we have 
\begin{equation}
\Ale^{(-)}_N \:\left( \Phi^{\mm}(z)\otimes\o_{\mm} \right) \;\in \;F^{\mm}, 
\end{equation}
and  comparing the $\gln$-weights of the vectors (\ref{eq:APhio}) and $\Omega^{(-)}_{\mm}$ we find that these vectors are proportional:
\begin{equation}
\Ale^{(-)}_N \:\left( \Phi^{\mm}(z)\otimes\o_{\mm} \right) = c(\mm)\,\Omega^{(-)}_{\mm} \qquad (\;c(\mm) \in \real\;).  \label{eq:APhi=cO}
\end{equation}

Now we observe that from the self-adjointness of the elementary permutations $K_{i,i+1}^{\dagger} = K_{i,i+1}$, $P_{i,i+1}^{\dagger} = P_{i,i+1}$ with respect to the scalar product (\ref{eq:sp}) it follows that the antisymmetrization operator is self-adjoint as well: 
\begin{equation}
{\Ale^{(-)}}^\dagger_N = \Ale^{(-)}_N.
\end{equation}
Therefore we can write
\begin{gather*}
\sprod{\;\Ale^{(-)}_N \:\left( \Phi^{\mm}(z)\otimes\o_{\mm} \right)\;}{\;\Ale^{(-)}_N \:\left( \Phi^{\mm}(z)\otimes\o_{\mm} \right)\;}  =   N!\: \sprod{\;\Phi^{\mm}(z)\otimes\o_{\mm}\;}{\;\Ale^{(-)}_N \:\left( \Phi^{\mm}(z)\otimes\o_{\mm} \right)\;} =  \\  =  N!\: c(\mm)\:\sprod{\;\Phi^{\mm}(z)\otimes\o_{\mm}\;}{\;\Omega^{(-)}_{\mm}\;}, 
\end{gather*}
and by the formula (\ref{eq:hwvector}) and the orthogonality of the polynomials $\Phi_{\s}^{\mm}(z)$ with respect to the scalar product (\ref{eq:spc}):
\begin{gather}
\sprod{\;\Ale^{(-)}_N \:\left( \Phi^{\mm}(z)\otimes\o_{\mm} \right)\;}{\;\Ale^{(-)}_N \:\left( \Phi^{\mm}(z)\otimes\o_{\mm} \right)\;} = c(\mm)^2\: {\sprod{\;\Omega^{(-)}_{\mm}\;}{\;\Omega^{(-)}_{\mm}\;}}_{(-)} = \\
=  N!\: c(\mm)\: {\sprod{\:\o_{\mm}\:}{\:\o_{\mm}\:}}_s \: {\sprod{\: \Phi^{\mm}(z)\:}{\: \Phi^{\mm}(z)\:}}_c.
\end{gather}
Using the definitions (\ref{eq:ol}, \ref{eq:baromegap}) to compute the norm ${\sprod{\:\o_{\mm}\:}{\:\o_{\mm}\:}}_s$ we  get: 
\begin{equation}
{\sprod{\;\Omega^{(-)}_{\mm}\;}{\;\Omega^{(-)}_{\mm}\;}}_{(-)} = N!\left(\prod_{s=1}^M p_s!\right)\frac{1}{c(\mm)} {\sprod{\: \Phi^{\mm}(z)\:}{\: \Phi^{\mm}(z)\:}}_c.
\end{equation}
The norms ${\sprod{\: \Phi^{\mm}(z)\:}{\: \Phi^{\mm}(z)\:}}_c$ are known, and can be found in the works \cite{Opdam,Cherednik2,Macdonald1}. For completeness we will give a  derivation of these norms later in this section. For now we will proceed to compute the coefficient $c(\mm)$.

Writing 
\begin{align}
& \Ale^{(-)}_N \:\left( \Phi^{\mm}(z)\otimes\o_{\mm} \right) = \sum_{\s \in S^{\mm}} \Phi_{\s}^{\mm}(z)\otimes \psi_{\s} \qquad (  \psi_{\s} \in \Ncplxn ) \\ 
\intertext{and (\ref{eq:hwvector})}  
& \Omega^{(-)}_{\mm} = \sum_{\s \in S^{\mm}} \Phi_{\s}^{\mm}(z)\otimes \check{\real}^{(-)}(\s)\o_{\mm} \\
\intertext{ from (\ref{eq:APhi=cO}) we get} 
& \psi_{\s} = c(\mm)\: \check{\real}^{(-)}(\s)\o_{\mm} \qquad ( \s \in S^{\mm} ). \label{eq:psiR}
\end{align}
Let $\bar{\s} \in S^{\mm}$ be the unique element of maximal length $l(\s)$ (\ref{eq:p2}) in the set $S^{\mm}$. This element corresponds to the anti-dominant rearrangement of the parts in the partition $\mm$:   
\begin{equation}
 \mlett_{\bar{\s}(1)} \leq \mlett_{\bar{\s}(2)} \leq \cdots \leq \mlett_{\bar{\s}(N)} . \label{eq:sigmabar}
\end{equation}
We will find the  coefficient $c(\mm)$  from (\ref{eq:psiR}) by comparing the vectors $\psi_{\bar{\s}}$ and $\check{\real}^{(-)}(\bar{\s})\o_{\mm}$.

First we compute the $\psi_{\bar{\s}}$. Let $\Sgroup{N}^{\mm}$ $\subset$ $\Sgroup{N}$ be the subgroup preserving the partition $\mm$. Then  
\begin{align}
&\Ale^{(-)}_N \:\left( \Phi^{\mm}(z)\otimes\o_{\mm} \right) = \sum_{\s \in S^{\mm}}(-1)^{l(\s)} K_{\s}P_{\s} \sum_{\tau \in \Sgroup{N}^{\mm}}(-1)^{l(\tau)} K_{\tau}P_{\tau} \:\Phi^{\mm}(z)\otimes\o_{\mm} \\
\intertext{and from (\ref{eq:Kphi=phi}), (\ref{eq:ol}, \ref{eq:baromegap})} 
&\Ale^{(-)}_N \:\left( \Phi^{\mm}(z)\otimes\o_{\mm} \right) = \left(\prod_{s=1}^Mp_s!\right) \sum_{\s \in S^{\mm}}(-1)^{l(\s)} K_{\s}P_{\s}\:\Phi^{\mm}(z)\otimes\o_{\mm}. \label{eq:intermed1}
\end{align}

\begin{lemma} For any element $\s \in S^{\mm}$ we have
\begin{align}
& K_{\s}\Phi^{\mm}(z) = \kappa^{\mm}(\s)\Phi_{\s}^{\mm}(z)  + \sum_{\s^{\prime} \in S^{\mm} \:\text{{\em s.t.}}\: l(\s^{\prime}) < l(\s)} \nu^{\mm}(\s,\s^{\prime})\Phi_{\s^{\prime}}^{\mm}(z) \label{eq:Li}\\
\intertext{where $\nu^{\mm}(\s,\s^{\prime}),  \kappa^{\mm}(\s)$ $\in$ $\real$ and {\em (\ref{eq:KactionPhi2})}}
& \kappa^{\mm}(\s(i,i+1)) = {\cal B}_i^{\mm}(\s) \kappa^{\mm}(\s) \qquad ( \mlett_{\s(i)} > \mlett_{\s(i+1)}). \label{eq:Lii}
\end{align} \label{l:lemma1}
\end{lemma}
\begin{pf} We prove the lemma by induction in the length of elements in $S^{\mm}$. For $\s ={\id}$ the (\ref{eq:Li}) trivially holds with $\kappa^{\mm}({\id}) = 1$. Fix a $\s \in S^{\mm}$ and assume that (\ref{eq:Li}) is true for all elements of the length less or equal to $l(\s)-1$. Then by the property (\ref{eq:p3}) there exists $(i,i+1)$ and $\tilde{\s} \in S^{\mm}$ such that $ \s = \tilde{\s}(i,i+1)$, $\mlett_{\tilde{\s}(i)} > \mlett_{\tilde{\s}(i+1)}$ and $ l(\tilde{\s}) = l(\s) - 1 $. Writing  
\begin{equation}
K_{\s}\Phi^{\mm}(z) = K_{i,i+1}K_{\tilde{\s}}\Phi^{\mm}(z)
\end{equation}
by the inductive assumption and (\ref{eq:KactionPhi1}) we obtain the desired statement. \end{pf}

Since the $\bar{\s}$ is the element of maximal length in $S^{\mm}$ from this lemma and  (\ref{eq:intermed1}) we find
\begin{align}
&\psi_{\bar{\s}} = (-1)^{l(\bar{\s})}\left(\prod_{s=1}^Mp_s!\right)\: \kappa^{\mm}(\bar{\s})\: P_{\bar{\s}}\o_{\mm} = (-1)^{l(\bar{\s})}\left(\prod_{s=1}^Mp_s!\right)\: \kappa^{\mm}(\bar{\s})\: \overline{\o}_{\mm} \\
\intertext{where}
& \overline{\o}_{\mm} =  P_{\bar{\s}}\:{\o}_{p_1}\otimes {\o}_{p_{2}}\otimes \cdots \otimes {\o}_{p_M}  = {\o}_{p_M}\otimes {\o}_{p_{M-1}}\otimes \cdots \otimes {\o}_{p_1}.
\end{align}
Now solving the recursion relation (\ref{eq:Lii}) with the initial condition $
\kappa^{\mm}({\id}) = 1$   we obtain
\begin{equation}
\kappa^{\mm}(\bar{\s}) = \prod\begin{Sb} 1\leq i < j \leq N \\ \mlett_i > \mlett_j \end{Sb} \frac{(\xi_i^{\mm} - \xi_j^{\mm})^2 - 1}{(\xi_i^{\mm} - \xi_j^{\mm})^2} \; = \; \prod_{1 \leq s < t \leq M} \frac{(h_{\mm}^{(s)}-h_{\mm}^{(t)} - p_s)(h_{\mm}^{(s)}-h_{\mm}^{(t)} + p_t)}{(h_{\mm}^{(s)}-h_{\mm}^{(t)})(h_{\mm}^{(s)}-h_{\mm}^{(t)} + p_t - p_s)}. \label{eq:kap}
\end{equation}

On the other hand using the recursion relation (\ref{eq:Rrecursion}) we get  
\begin{equation}
\check{\real}^{(-)}(\bar{\s})\o_{\mm} = (-1)^{l(\bar{\s})}\left(\prod_{1\leq s < t\leq M}a_{s,t}\left(h_{\mm}^{(s)}-h_{\mm}^{(t)}\right)\right)\: \overline{\o}_{\mm} 
\end{equation}
where 
\begin{equation}
a_{s,t}(x) := \begin{cases}\; \frac{x + p_t}{x + p_t - p_s}  & \;  (\; p_s \leq p_t \;),\\
\; \frac{x + p_t}{x}  & \;  (\; p_s \geq p_t \;). \end{cases} \label{eq:aa}
\end{equation}
Hence introducing  
\begin{align}
& \rho(\mm) := \left(\prod_{1\leq s < t\leq M}\rho_{s,t}\left(h_{\mm}^{(s)}-h_{\mm}^{(t)}\right)\right), \\
& \rho_{s,t}(x) :=  \begin{cases}\; \frac{x}{x - p_s}  & \;  (\; p_s \leq p_t \;),\\
\; \frac{x + p_t-p_s}{x - p_s}  & \;  (\; p_s \geq p_t \;). \end{cases}
\end{align}
we find from (\ref{eq:kap}) and (\ref{eq:aa}) that
\begin{equation}
c(\mm) = \left(\prod_{s=1}^Mp_s!\right) \frac{1}{\rho(\mm)}
\end{equation}
and 
\begin{equation}
{\sprod{\:\Omega^{(-)}_{\mm}\:}{\:\Omega^{(-)}_{\mm}\:}}_{(-)} = N!\:\rho(\mm)\:{\sprod{\:\Phi^{\mm}(z)\:}{\:\Phi^{\mm}(z)\:} }_c. \label{eq:omnorm1}
\end{equation}

\begin{prop}
For $\mm \in \MC_N$ we have $(\xi_i^{\mm}:= \alpha\mlett_i - i)$:
\begin{equation}
{\sprod{\:\Phi^{\mm}(z)\:}{\:\Phi^{\mm}(z)\:} }_c = \frac{1}{N!}\prod_{1\leq i < j \leq N} \frac{\Gamma\left(\:\frac{\xi_i^{\mm}-\xi_j^{\mm}}{\alpha} + \frac{1}{\alpha} + 1\:\right) \: \Gamma\left(\:\frac{\xi_i^{\mm}-\xi_j^{\mm}}{\alpha} - \frac{1}{\alpha} + 1\: \right)}{\left\{\Gamma\left(\:\frac{\xi_i^{\mm}-\xi_j^{\mm}}{\alpha}  + 1 \:\right)\right\}^2},  \label{eq:Phinorm1}
\end{equation}
or, equivalently,  in notations {\em (\ref{eq:p_s}, \ref{eq:h^s})}:
\begin{multline}
{\sprod{\:\Phi^{\mm}(z)\:}{\:\Phi^{\mm}(z)\:} }_c = \\ = \frac{1}{N!}\left(\prod_{s=1}^M\frac{\Gamma\left(\:\frac{p_s}{\alpha}+1\:\right)}{\left\{\Gamma\left(\:\frac{1}{\alpha}+1\:\right)\right\}^{p_s}}\right)\:\prod_{1\leq s < t \leq M}\frac{\Gamma\left(\:\frac{h_{\mm}^{(s)}-h_{\mm}^{(t)}}{\alpha} + \frac{p_t}{\alpha} + 1\:\right) \:\Gamma\left(\:\frac{h_{\mm}^{(s)}-h_{\mm}^{(t)}}{\alpha} - \frac{p_s}{\alpha} + 1\:\right)}{\Gamma\left(\:\frac{h_{\mm}^{(s)}-h_{\mm}^{(t)}}{\alpha} + \frac{p_t-p_s}{\alpha} + 1\:\right)\:\Gamma\left(\:\frac{h_{\mm}^{(s)}-h_{\mm}^{(t)}}{\alpha} +  1\:\right)}.\label{eq:Phinorm2}
\end{multline} \label{p:prop3}
\end{prop}
\begin{pf} To prove the proposition we will use the known formula for the norms of symmetric Jack polynomials. The Jack polynomial $P_{\mm}^{(\alpha)}(z)$ \cite{Macdonaldbook} is the unique symmetric vector in the space $ E^{\mm} := \oplus_{\s\in S^{\mm}}\cplx \Phi_{\s}^{\mm}(z)$ $(\mm \in \MC_N)$ normalized so that in the expansion
\begin{equation}
P_{\mm}^{(\alpha)}(z) = \sum_{\s \in S^{\mm}} \nu^{\mm}(\s)\: \Phi_{\s}^{\mm}(z) \qquad (\nu^{\mm}(\s) \in \real) \label{eq:Jack}
\end{equation}
the coefficient $\nu^{\mm}({\id})$ is equal to 1.
The symmetry conditions 
\begin{equation}
K_{i,i+1} P_{\mm}^{(\alpha)}(z) = P_{\mm}^{(\alpha)}(z) \qquad (i=1,2,\dots,N-1)
\end{equation}
together with the formulas (\ref{eq:KactionPhi1}, \ref{eq:KactionPhi2}) give the recursion relation 
\begin{equation}
\nu^{\mm}(\s(i,i+1)) = ( 1 - {\cal A}_i(\s))\nu^{\mm}(\s) = \frac{ \xi_i^{\mm}(\s)- \xi_{i+1}^{\mm}(\s)  - 1}{ \xi_i^{\mm}(\s)- \xi_{i+1}^{\mm}(\s)}\nu^{\mm}(\s) \qquad (\mlett_{\s(i)} > \mlett_{\s(i+1)}).
\end{equation}
Solving this relation with the initial condition $\nu^{\mm}({\id})=1$ gives  
\begin{equation}
\nu^{\mm}(\bar{\s}) = \prod\begin{Sb} 1\leq i < j \leq N \\ \mlett_i > \mlett_j \end{Sb} \frac{\xi^{\mm}_i - \xi^{\mm}_j - 1}{\xi^{\mm}_i - \xi^{\mm}_j}
\end{equation}
where $\bar{\s}$ is the element of maximal length in the set $S^{\mm}$ (\ref{eq:sigmabar}).

Let $\text{Symm}_N$ $:=$ $\sum_{\s\in \Sgroup{N}} K_{\s} $ be the symmetrization operator in $\cz$. Then 
\begin{equation}
\text{Symm}_N \Phi^{\mm}(z) = d(\mm) P_{\mm}^{(\alpha)}(z). \qquad ( d(\mm) \in \real )\label{eq:SymPhi-Jack}
\end{equation}
Writing 
\begin{equation}
\text{Symm}_N \Phi^{\mm}(z) = \sum_{\s\in S^{\mm}} K_{\s} \sum_{\tau\in \Sgroup{N}^{\mm}} K_{\tau} \Phi^{\mm}(z)  
\end{equation}
and using (\ref{eq:Kphi=phi}) and the result of the lemma \ref{l:lemma1} we get\begin{equation} 
\text{Symm}_N \Phi^{\mm}(z) = \left(\prod_{s=1}^Mp_s!\right)\kappa^{\mm}(\bar{\s}) \Phi_{\bar{\s}}^{\mm}(z) + \sum\begin{Sb} \s \in S^{\mm} \\ \s \neq \bar{\s} 
\end{Sb} \zeta(\s) \: \Phi_{\s}^{\mm}(z) \qquad  (\zeta(\s) \in \real)
\end{equation}
where $\kappa^{\mm}(\bar{\s})$ is given by (\ref{eq:kap}). Comparing the last equation with (\ref{eq:Jack}) gives for the coefficient $d(\mm)$ in (\ref{eq:SymPhi-Jack}): 
\begin{equation}
d(\mm) = \left(\prod_{s=1}^Mp_s!\right) \: \frac{\kappa^{\mm}(\bar{\s})}{\nu^{\mm}(\bar{\s})} = \prod_{1\leq i < j \leq N}\frac{\xi^{\mm}_i - \xi^{\mm}_j + 1}{\xi^{\mm}_i - \xi^{\mm}_j}.  \label{eq:c(lambda)}
\end{equation}

Now the self-adjointness of the symmetrization operator with respect to the scalar product (\ref{eq:spc}) yields  
\begin{equation}
{\sprod{\:\Phi^{\mm}(z)\:}{\:\Phi^{\mm}(z)\:}}_c = \frac{1}{N!}\,d(\mm)\, {\:\sprod{P_{\mm}^{(\alpha)}(z)\:}{P_{\mm}^{(\alpha)}(z)\:}}_c.
\end{equation}

Using the expression \cite[Ch.VI-10.38]{Macdonaldbook}: 
\begin{equation}
{\:\sprod{P_{\mm}^{(\alpha)}(z)\:}{P_{\mm}^{(\alpha)}(z)\:}}_c = \prod_{1\leq i < j \leq N} \frac{\Gamma\left(\:\frac{\xi_i^{\mm}-\xi_j^{\mm}}{\alpha} + \frac{1}{\alpha} \:\right) \: \Gamma\left(\:\frac{\xi_i^{\mm}-\xi_j^{\mm}}{\alpha} - \frac{1}{\alpha} + 1\: \right)}{\Gamma\left(\:\frac{\xi_i^{\mm}-\xi_j^{\mm}}{\alpha}  + 1 \:\right)\:\Gamma\left(\:\frac{\xi_i^{\mm}-\xi_j^{\mm}}{\alpha} \:\right)} \qquad (\xi_i^{\mm}:= \alpha\mlett_i - i) \label{eq:Jacknorm}
\end{equation}
and (\ref{eq:c(lambda)}) we get (\ref{eq:Phinorm1}). The formula (\ref{eq:Phinorm2}) follows from (\ref{eq:Phinorm1}) by using the notations (\ref{eq:p_s}) and (\ref{eq:h^s}).
\end{pf}

Now combining the result of the proposition \ref{p:prop3} and the formula (\ref{eq:omnorm1}) together we obtain the main result of this subsection: 

\begin{prop}
For $\mm \in \MC^{(n)}_N$ we have 
\begin{multline}
{\sprod{\:\Omega^{(-)}_{\mm}\:}{\:\Omega^{(-)}_{\mm}\:} }_{(-)} = \label{eq:Omeganorm2} \\  \prod_{s=1}^M\frac{\Gamma\left(\:\frac{p_s}{\alpha}+1\:\right)}{\left\{\Gamma\left(\:\frac{1}{\alpha}+1\:\right)\right\}^{p_s}}\:\prod_{1\leq s < t \leq M}\frac{\Gamma\left(\:\frac{h_{\mm}^{(s)}-h_{\mm}^{(t)}}{\alpha} + \frac{p_t}{\alpha} + 1\:\right) \:\Gamma\left(\:\frac{h_{\mm}^{(s)}-h_{\mm}^{(t)}}{\alpha} - \frac{p_s}{\alpha} \:\right)}{\Gamma\left(\:\frac{h_{\mm}^{(s)}-h_{\mm}^{(t)}}{\alpha} + \frac{p_t-p_s}{\alpha} + \theta(p_s\leq p_t)\:\right)\:\Gamma\left(\:\frac{h_{\mm}^{(s)}-h_{\mm}^{(t)}}{\alpha} +  \theta(p_s > p_t)\:\right)} \nonumber 
\end{multline} 
where 
\begin{equation}
\theta(x):= \begin{cases} \; 1 \; &  \text{{\em when $x$ is true,} } \\ \; 0 \; &  \text{{\em when $x$ is false.} } \end{cases}
\end{equation}
\label{p:prop4}
\end{prop}
\subsection{The norm of the ground state in the Fermionic case}
The ground state $\Omega^{(-)}_{\mm^0(N)}$ of the fermionic SCSM is identified with the highest-weight vector in the Yangian sub-representation $F^{\mm^0(N)}$ where the ground-state partition $\mm^0(N)$ is described as follows. For a given number of particles $N$ let $L \in \{0\}\cup{\Bbb N}$ and $q \in \{0,1,\dots,n-1\}$ be defined by $N = n L + q$. Then\begin{equation}    
\mm^0(N) = \begin{cases} (L)^n (L-1)^n \cdots (1)^n (0)^q  & \quad ( q \neq 0 ),\\ (L-1)^n (L-2)^n \cdots (1)^n (0)^n  & \quad ( q = 0 )
\end{cases}
\end{equation}
where we used the usual convention: $ (a)^r = \underbrace{a,a,\dots,a }_r$.
The  ground state $\Omega^{(-)}_{\mm^0(N)}$ has degeneracy equal to $\dim F^{\mm^0(N)}$ = $ \left(\begin{array}{c} n \\ q \end{array}\right)$. 

One expression for $\Omega^{(-)}_{\mm_0(N)}$  is given by the formula (\ref{eq:hwvector}). For the special case of the ground state partition $\mm^0(N)$ this expression can be simplified by taking into account the triangularity of the non-symmetric Jack polynomials (\ref{eq:triangularity}). This gives  
\begin{equation}
\Omega^{(-)}_{\mm^0(N)} = \Ale_N^{(-)}\left( z_1^{\mlett^0(N)}z_2^{\mlett^0(N)}\cdots  z_N^{\mlett^0(N) } \otimes \left( (v_1\otimes v_2\otimes \cdots \otimes v_n)^{\otimes L} \right) \otimes v_1 \otimes v_2 \otimes \cdots \otimes v_q \right).
\end{equation}
Let us introduce the Laurent polynomials $ \tilde{f}_{\mm^0(N)}(z_1,z_2,\dots,z_N) $ and $ f_{\mm^0(N)}(z_1,z_2,\dots,z_N) $ by  
\begin{align*}
& \tilde{f}_{\mm^0(N)}(z_1,z_2,\dots,z_N) := \prod_{\ep = 1}^q \left(\prod_{ (\ep - 1)(L+1) < i < j \leq \ep (L+1) } (z_i - z_j) \right) \prod_{\ep = q+1}^n \left( \prod_{ q+(\ep - 1)L < i < j \leq q+\ep L } (z_i - z_j)  \right) \\
\intertext{and}
& f_{\mm^0(N)}(z_1,z_2,\dots,z_N) := \begin{cases} \left(\prod_{i=q(L+1)+1}^N z_i \right) \tilde{f}_{\mm^0(N)}(z_1,z_2,\dots,z_N) & \quad (q \neq 0), \\ \tilde{f}_{\mm^0(N)}(z_1,z_2,\dots,z_N) & \quad (q = 0). \end{cases}
\end{align*}
And let the sequence  $(\ep^0_1,\ep^0_2,\dots \ep^0_N)$ be defined as follows
\begin{equation}
(\ep^0_1,\ep^0_2,\dots \ep^0_N) := \begin{cases} (1)^{L+1} (2)^{L+1}\cdots (q)^{L+1}(q+1)^L \cdots (n)^L & \quad (q \neq 0) ,\\  (1)^L (2)^L \cdots (n)^L & \quad (q=0).\end{cases}
\end{equation}
Then up to a sign the ground state $\Omega^{(-)}_{\mm^0(N)}$ can be represented as 
\begin{equation}
\sum_{\s } (-1)^{l(\s)}f_{\mm^0(N)}(z_{\s(1)},z_{\s(2)},\dots,z_{\s(N)}) \otimes \left( v_{\ep^0_{\s(1)}} \otimes v_{\ep^0_{\s(2)}} \otimes \cdots \otimes v_{\ep^0_{\s(N)}} \right)
\end{equation}
where the sum is taken over all permutations such that the corresponding sequences \\ $ (\ep^0_{\s(1)},\ep^0_{\s(2)},\dots \ep^0_{\s(N)}) $ are all distinct.

Using this presentation we can write the norm of the ground state as 
\begin{equation}
{\sprod{\:\Omega^{(-)}_{\mm^0(N)}\:}{\:\Omega^{(-)}_{\mm^0(N)}\:}}_{(-)} = \frac{N!}{\left\{(L+1)!\right\}^q \left\{L!\right\}^{n-q}}\,{\sprod{\:f_{\mm^0(N)}(z_1,z_2,\dots,z_N)\:}{\:f_{\mm^0(N)}(z_1,z_2,\dots,z_N)\:} }_c.  
\end{equation}
By the definition (\ref{eq:spc}) of the scalar product on the space of Laurent polynomials we can now recast the statement of the proposition \ref{p:prop4} for the case of the ground state as the following integral formula   
\begin{gather}  
{\sprod{\:\Omega^{(-)}_{\mm^0(N)}\:}{\:\Omega^{(-)}_{\mm^0(N)}\:}}_{(-)} = \\\frac{1}{\left\{(L+1)!\right\}^q \left\{L!\right\}^{n-q}}\:\left( \prod_{i=1}^N\oint _{|w_i| = 1}\frac{dw_i}{2\pi \sqrt{-1}w_i}  \right) \prod_{ i < j}|w_i - {w_j} |^{\frac{2}{\alpha}}  |f_{\mm^0(N)}(w_1,w_2,\dots,w_N)|^2 = \nonumber \\
= \frac{\Gamma\left( (\frac{n}{\alpha}+1)L + \frac{q}{\alpha} + 1 \right)}{ L! (\frac{n}{\alpha}+1)^L \Gamma\left( \frac{1}{\alpha} + 1 \right)^N }.\nonumber \label{eq:grstsp} 
\end{gather}

\subsection{Bosonic case}

The computation of the norms of the Yangian highest weight vectors in the bosonic case is much  simpler than that in  the fermionic case. \mbox{} From the equation (\ref{eq:hwvector+}) and the definition of the Jack polynomial we immediately find\begin{equation}   
\Omega_{\mm}^{(+)} =  P^{(\alpha)}_{\mm}(z) \otimes ( v_1^{\otimes N}) \qquad (\mm \in \MC_N ).
\end{equation}
Hence the norm of the highest weight vector is given by
\begin{equation}
{\sprod{\:\Omega^{(+)}_{\mm}\:}{\:\Omega^{(+)}_{\mm}\:}}_{(+)} = {\sprod{\:P_{\mm}^{(\alpha)}(z)\:}{\:P_{\mm}^{(\alpha)}(z)\:}}_c  \label{eq:boshwnorm}
\end{equation}
where the norm ${\sprod{\:P_{\mm}^{(\alpha)}(z)\:}{\:P_{\mm}^{(\alpha)}(z)\:}}_c $ of the Jack polynomial is given by the formula (\ref{eq:Jacknorm}). 


\section{Eigenbases of the Gelfand-Zetlin algebra in the irreducible Yangian submodules and norms of the eigenvectors} \label{sec:dnorms}
In this section we construct eigenbases of the operator-valued series $A_1^{(\kappa)}(u), A_2^{(\kappa)}(u), \dots A_n^{(\kappa)}(u)$ within each of the irreducible $Y(\gln)$-submodules $F^{\mm}$ $(\mm \in \MC^{(n)}_N)$ ( $\kappa = -1$ -- fermionic case ) and $B^{\mm}$ $(\mm \in \MC_N)$ ( $\kappa = 1$ -- bosonic case ), and compute norms of the eigenvectors that form these eigenbases.
 
Due to the isomorphisms given by the theorems \ref{t:FDEC} and \ref{t:BDEC} the construction of the eigenbases is carried out by a straightforward application of the results of Nazarov and Tarasov that are summarized in sec. \ref{sec:background}.

Let us fix a partition $\mm = (\mlett_1,\mlett_2,\dots,\mlett_N)  \in \MC_N$ and let for $\kappa=-1$ $\mm \in \MC^{(n)}_N \subset \MC_N$. As in sec. \ref{sec:decomposition} associate with the $\mm$ the following data: \\
$M$ -- the number of distinct elements in the sequence $\mm = (\mlett_1,\mlett_2,\dots,\mlett_N)$;  $ p_s $ $( s=1,2,\dots,M)$ -- the multiplicities of the elements in the $\mm$:   
\begin{multline}
\mlett_1=\mlett_2=\cdots =\mlett_{p_1} > \mlett_{1+p_1}= \mlett_{2+p_1}=\cdots =\mlett_{p_2+p_1}> \quad \cdots \\ \cdots > \mlett_{1+ p_{M-1}+\cdots +p_2+p_1}=  \mlett_{2+ p_{M-1}+\cdots +p_2+p_1}=\cdots =\mlett_{p_{M}+\cdots +p_2+p_1 \equiv N} .  \label{eq:p_s1}
\end{multline}
Since in the fermionic case the partition is restricted : $\mm \in \MC^{(n)}_N$, we have $  p_s \in \{1,2,\dots,n\} $ $( s=1,2,\dots,M)$ when $\kappa =-1$. 

With $\xi^{\mm}_i := \xi^{\mm}_i({\id})= \alpha \mlett_i - i $ (\ref{eq:d-eigenvectors}) set 
\begin{equation}
h_{\mm}^{(s)} := -\kappa \xi^{\mm}_{1+p_1+p_2+\cdots+p_{s-1}} \qquad (p_0 :=0,\quad s=1,2,\dots,M ). \label{eq:h^s1}
\end{equation}

For $ p\in \{1,2,\dots,n\} $ let $ {\cal S}^{(-)}_p $ denote the set of all Gelfand-Zetlin schemes $ \Lambda $ that are associated with the irreducible $\gln$-module with  the highest weight (cf. sec. \ref{sec:background})
\begin{equation}
 (\underbrace{1,1,\dots,1}_{p},\underbrace{0,0,\dots,0}_{n-p}).
\end{equation}
An element of $ {\cal S}^{(-)}_p $ is an array of the form
\begin{eqnarray}
& \l _{n,1} \; \; \l _{n,2} \; \cdots \cdots \cdots \cdots \cdots \; \; 
 \l _{n,n} &  \\
& \l _{n-1,1} \; \;   \cdots \cdots \; \; \l _{n-1,n-1} &
\nonumber \\
& \ddots \; \; \cdots \cdots \; \; \; \; \; \; \;  & \nonumber \\
& \l _{2,1} \; \; \l _{2,2} & \nonumber \\
& \l _{1,1} & \nonumber     \label{eq:gzscheme}
\end{eqnarray}
where 
\begin{gather}
(\l_{m,1},\l_{m,2}, \dots ,\l_{m,m}) = (\underbrace{1,1,\dots,1}_{l_{m}},\underbrace{0,0,\dots,0}_{m-l_{m}}) \qquad ( m=1,2,\dots,n), \label{eq:gzferm} \\ 
l_n = p \nonumber  
\end{gather}
and 
\begin{equation}
\text{either $ \quad l_{m} = l_{m+1} \quad $  or  $  \quad l_{m} = l_{m+1}-1 \quad $ } \qquad ( m=1,2,\dots,n-1).
\end{equation}

For $ p\in {\Bbb N} $ let $ {\cal S}^{(+)}_p $ denote the set of all Gelfand-Zetlin schemes $ \Lambda = \backslash \lambda_{m,m'} /_{n\geq m\geq m'\geq 1}$ that are associated with the irreducible $\gln$-module with the highest weight (cf. sec. \ref{sec:background})
\begin{equation}
 ({p},\underbrace{0,0,\dots,0}_{n-1}).
\end{equation}
An element of  $ {\cal S}^{(+)}_p $ is a Gelfand-Zetlin scheme of the form 
\begin{eqnarray}
& \alpha _{n} \; \; 0  \; \cdots \cdots \cdots \cdots \cdots \; \; 
 0 &  \label{eq:gzbos} \\
& \alpha _{n-1} \; \;  0 \; \;  \cdots \cdots \; \; 0 &
\nonumber \\
& \ddots \; \; \cdots \cdots \; \; \; \; \; \; \;  & \nonumber \\
& \alpha _{2} \; \; 0 & \nonumber \\
& \; \; \alpha _{1} & \nonumber 
\end{eqnarray}
where 
\begin{gather}
\alpha_{m} \leq \alpha_{m+1} \qquad ( m=1,2,\dots,n-1), \\
\alpha_n = p. \nonumber  
\end{gather}

Now let us define the following operator-valued series:\\
For the bosonic case  set
\begin{equation}
a_{m}^{(+)}(u)= A^{(+)}_m(u), \; \;  b_{m}^{(+)}(u)= B^{(+)}_m(u), \; \; 
c_{m}^{(+)}(u)=  C^{(+)}_m(u), \; \; , d_{m}^{(+)}(u)= D^{(+)}_m(u). 
\end{equation}
And for the fermionic case   set
\begin{gather}
a_{m}^{(-)}(u)= \Delta (u) A^{(-)}_m(u), \; \;  
b_{m}^{(-)}(u)= \Delta (u) B^{(-)}_m(u), \; \;  \\
c_{m}^{(-)}(u)= \Delta (u) C^{(-)}_m(u), \; \; 
d_{m}^{(-)}(u)= \Delta (u) D^{(-)}_m(u) \nonumber 
\end{gather}
where $\Delta (u) = \prod _{i=1}^{N} (u+d_i)$.
Then from the proposition \ref{p:sa} it follows that 
\begin{equation} \label{adj}
a_{m}^{(\kappa)}(u) ^{\dagger }= a_m^{(\kappa)} (u), \; \; \; 
b_{m}^{(\kappa)}(u) ^{\dagger }= c_m ^{(\kappa)}(u)  , \; \; \;  
c_{m}^{(\kappa)}(u) ^{\dagger }= b_m ^{(\kappa)}(u),  \; \; \; \kappa=-,+ .\label{con}
\end{equation} 

For a collection of Gelfand-Zetlin schemes $\Lambda ^{(1)}, \dots ,\Lambda ^{(M)}$ such that $ \Lambda ^{(s)} \in {\cal S}^{(\kappa)}_{p_s} $ $ (s=1,2,\dots,M)$ define the following vector (cf. sec. \ref{sec:background}):
\begin{align}
& v^{(\kappa)}_{ \Lambda ^{(1)}, \dots ,\Lambda ^{(M)}}  =
\prod_{(m,m')}^\rightarrow\
\left( \prod_{(s,t) \atop{1 \leq t \leq \l ^{(s)}_{n,m'}- \l ^{(s)}_{m,m'}}}
b_m^{(\kappa)}({\nu }_{m,m'}^{(s)}-t)
\right) \cdot \Omega^{(\kappa)}_{\mm}, \\
& v^{(\kappa)}_{ \Lambda ^{(1)}, \dots ,\Lambda ^{(M)}} \in \begin{cases} F^{\mm} 
 &  (\kappa = - ), \\ B^{\mm} &  (\kappa = + ). \end{cases} 
\end{align}
Here 
\begin{equation}
{\nu }_{m,m'}^{(s)} = m'-\l _{m,m'}^{(s)}-1 - h^{(s)}_{\mm} \label{nu}
\end{equation}
and the $h^{(s)}_{\mm}$ are defined by (\ref{eq:h^s1}).
\mbox{} From  the proposition \ref{base} and the theorems \ref{t:FDEC},\ref{t:BDEC} it follows that  the set 
\begin{equation}
  \{ v^{(\kappa)}_{ \Lambda ^{(1)}, \dots ,\Lambda ^{(M)}} \; | \;  
 \; \Lambda ^{(s)}
 \in {\cal S}^{(\kappa)}_{p_s}\quad (s=1,2,\dots,M) \} \label{eq:base}
\end{equation}
is a base of  $F^{\mm }$  ( resp. $B^{\mm}$) when $ \kappa = - $ ( resp. $ +$). Due to the proposition \ref{am} this is an eigenbase of the operators generating the Gelfand-Zetlin algebra:  
\begin{align}
& A_m^{(\kappa)}(u)\, v^{(\kappa)}_{ \Lambda ^{(1)}, \dots ,\Lambda ^{(M)}} = {\cal A}_m^{(\kappa)}(u;\mm)_{ \Lambda ^{(1)}, \dots ,\Lambda ^{(M)}} v^{(\kappa)}_{ \Lambda ^{(1)}, \dots ,\Lambda ^{(M)}}, \qquad (m=1,2,\dots,n)\\
\intertext{where the eigenvalues are defined by :}
& {\cal A}_m^{(-)}(u;\mm)_{ \Lambda ^{(1)}, \dots ,\Lambda ^{(M)}} = \prod_{s=1}^M \frac{u + 1 + h^{(s)}_{\mm}} {u + 1 + h^{(s)}_{\mm} - l_m^{(s)}},  \qquad (\Lambda^{(s)} \in {\cal S}^{(-)}_{p_s});\label{eq:eigenferm}\\ 
&{\cal A}_m^{(+)}(u;\mm)_{ \Lambda ^{(1)}, \dots ,\Lambda ^{(M)}} = \prod_{s=1}^M \frac{u + h^{(s)}_{\mm} + \alpha_m^{(s)}} {u + h^{(s)}_{\mm}},\:\quad\qquad (\Lambda^{(s)} \in {\cal S}^{(+)}_{p_s}).\label{eq:eigenbose}
\end{align}

Since ${\sprod{\; \Phi ^{\mm }_{\s }(z)\;}{\; \Phi ^{\nn}_{\tau }(z)\;}}_c =0 $ when $ \mm \neq \nn $,  the subspaces $F^{\mm }$ (resp. $B^{\mm }$)
 are pairwise orthogonal. 

For $\alpha > 0$ one can verify, that the data $\mm \in \MC_N$, $( \Lambda ^{(1)},\Lambda ^{(2)} \dots ,\Lambda ^{(M)})$ $(\Lambda ^{(s)}  \in  {\cal S}^{(\kappa)}_{p_s})$ are uniquely restored from the collection of rational functions
\begin{equation}
{\cal A}_1^{(\kappa)}(u;\mm)_{ \Lambda ^{(1)}, \dots ,\Lambda ^{(M)}},{\cal A}_2^{(\kappa)}(u;\mm)_{ \Lambda ^{(1)}, \dots ,\Lambda ^{(M)}},\dots,{\cal A}_n^{(\kappa)}(u;\mm)_{ \Lambda ^{(1)}, \dots ,\Lambda ^{(M)}}.
\end{equation}
That is the joint spectrum  of eigenvalues of the Gelfand-Zetlin algebra is simple. Since $A_m ^{(\kappa)}(u)$ are self-adjoint , we obtain 
\begin{prop}
For $\mm \in \MC_N^{(n)} $  ( resp.  $\mm \in \MC_N $ )  the set 
\begin{equation}
  \{ v^{(\kappa)}_{ \Lambda ^{(1)}, \dots ,\Lambda ^{(M)}} \; | \;  
 \; \Lambda ^{(s)}
 \in {\cal S}^{(\kappa)}_{p_s}\quad (s=1,2,\dots,M) \}
\end{equation}
with $\kappa = - $ ( resp. $\kappa = + $)  is an  orthogonal base of $F^{\mm }$  ( resp. $B^{\mm }$ ).
\end{prop}

The norms of the eigenvectors  $v^{(\kappa)}_{ \Lambda ^{(1)},\dots ,\Lambda ^{(M)}}$ are as follows
\begin{prop} \label{norm}

(Bosonic case) 

Let $\mm \in \MC_N$ and $ \Lambda^{(s)} \in {\cal S}^{(+)}_{p_s} $ $( s=1,2,\dots,M).$ If we write a Gelfand-Zetlin scheme $ \Lambda^{(s)}$ as in (\ref{eq:gzbos}): 
\begin{eqnarray}
& \alpha _{n}^{(s)} \; \; 0  \; \cdots \cdots \cdots \cdots \cdots \; \; 
 0 &  \\
& \alpha _{n-1}^{(s)} \; \;  0 \; \;  \cdots \cdots \; \; 0 &
\nonumber \\
\Lambda^{(s)} = & \ddots \; \; \cdots \cdots \; \; \; \; \; \; \;  & \nonumber \\
& \alpha _{2}^{(s)} \; \; 0 & \nonumber \\
& \; \; \alpha _{1}^{(s)} & \nonumber
\end{eqnarray}
then the norm of the vector $v^{(+)}_{ \Lambda ^{(1)},\dots ,\Lambda ^{(M)}}$ is 
 \begin{multline}
{\langle  v^{(+)}_{ \Lambda ^{(1)}, \dots ,\Lambda ^{(M)}},
v^{(+)}_{ \Lambda ^{(1)}, \dots ,\Lambda ^{(M)}} \rangle}_{(+)}
=  
\qquad  \qquad {\sprod{\:\Omega_{\mm}^{(+)}\:}{\:\Omega_{\mm}^{(+)}\:}}_{(+)}
 \cdot \\
\cdot  \prod _{1 \leq m \leq n}
\left\{
\prod _{1 \leq s \leq M}
\frac{(\alpha _{n}^{(s)} - \alpha _{m}^{(s)})!
(\alpha _{n}^{(s)} - \alpha _{m-1}^{(s)})!
(\alpha _{m}^{(s)}!)^2}{
(\alpha _{m}^{(s)} - \alpha _{m-1}^{(s)})!
(\alpha _{n}^{(s)}!)^2}
\right.
\nonumber  \\
\left\{ 
\prod _{(s,s') \atop{s \neq s'} } 
\prod _{a=\alpha _{m}^{(s)}}^{\alpha _{n}^{(s)} -1}
\frac{
(-a+\alpha _{n}^{(s')} +h_{\mm}^{(s')}-h_{\mm}^{(s)} )
(-1-a+\alpha _{m-1}^{(s')} +h_{\mm}^{(s')}-h_{\mm}^{(s)} )}{
(-1-a +h_{\mm}^{(s')}-h_{\mm}^{(s)} )^2}
\right\}
\nonumber \\
\left. 
\prod _{(s,s') \atop{s < s'} }
\frac{(\alpha _{n}^{(s')}-\alpha _{n}^{(s)} +h_{\mm}^{(s')}-h_{\mm}^{(s)})}{
(\alpha _{m}^{(s')}-\alpha _{m}^{(s)} +h_{\mm}^{(s')}-h_{\mm}^{(s)})}
\right\}
\nonumber \\  \mbox{}\nonumber 
\end{multline}


where the $h_{\mm}^{(s)}$ are defined by (\ref{eq:h^s1}) with $\kappa = +$.
\\ \mbox{} \\ 
(Fermionic case)

Let $\mm \in \MC_N^{(n)}$ and $ \Lambda^{(s)} \in {\cal S}^{(-)}_{p_s} $ $( s=1,2,\dots,M).$
As in (\ref{eq:gzferm}) define $l_m^{(s)}$ associated with the Gelfand-Zetlin scheme $\Lambda^{(s)}$ by the conditions $\l _{m,l_m^{(s)}}^{(s)} =1$ and
 $\l _{m,l_m^{(s)}+1}^{(s)} =0$. Then the norm of the vector $ v^{(-)}_{ \Lambda ^{(1)}, \dots ,\Lambda ^{(M)}}$ is 
\begin{multline}
{\langle  v^{(-)}_{ \Lambda ^{(1)}, \dots ,\Lambda ^{(M)}},
v^{(-)}_{ \Lambda ^{(1)}, \dots ,\Lambda ^{(M)}} \rangle}_{(-)}
= \qquad \qquad {\sprod{\:\Omega_{\mm}^{(-)}\:}{\:\Omega_{\mm}^{(-)}\:}}_{(-)} \cdot \\
\cdot \left\{
\prod _{1 \leq s \leq M}
\prod_{(m,m')  \atop{\lambda_{m,m'}^{(s)} \neq \lambda _{n,m'}^{(s)}}} 
(m'-1)!^{2}(p_s +1-m')!^{2}
\right\} \\ \mbox{}  \nonumber  
\end{multline}
\begin{equation}
\left\{
\prod _{(s,s') \atop{s < s'} }
\prod_{(m,m')  \atop{\lambda_{m,m'}^{(s)} \neq \l _{n,m'}^{(s)}
\atop{\lambda_{m,m'}^{(s')} \neq \l _{n,m'}^{(s')}}}} 
\frac{\prod_{j=0}^{p_s}(m'-j-1+h_{\mm}^{(s')}-h_{\mm}^{(s)})^{2}
\prod_{j=0}^{p_{s'}}(m'-j-1+h_{\mm}^{(s)}-h_{\mm}^{(s')})^{2}}{
(h_{\mm}^{(s)}-h_{\mm}^{(s')})^{4}}
\right\}
\nonumber
\end{equation}
\begin{equation}
\left\{
\prod _{(s,s') \atop{s \neq s'} }
\prod_{(m,m')  \atop{\lambda_{m,m'}^{(s)} \neq \l _{n,m'}^{(s)}
\atop{\lambda_{m,m'}^{(s')} = \l _{n,m'}^{(s')}}}} 
\frac{(m'-l_{m}^{(s')}+h_{\mm}^{(s')}-h_{\mm}^{(s)})
\prod_{j=0}^{p_{s}}(m'-j-1+h_{\mm}^{(s')}-h_{\mm}^{(s)})^{2}}{
(m'-1-l_{m-1}^{(s')}+h_{\mm}^{(s')}-h_{\mm}^{(s)})
(m'-1-l_{m}^{(s')}+h_{\mm}^{(s')}-h_{\mm}^{(s)})
(m'-l_{m+1}^{(s')}+h_{\mm}^{(s')}-h_{\mm}^{(s)})}
\right\}
\nonumber
\end{equation}
where $h_{\mm}^{(s)}$ are defined by (\ref{eq:h^s1}) with $\kappa = -$.
In these product formulas the $s$ and $s'$ range from $1$ to $M$ (\ref{eq:p_s1}) and $ (m,m')$ ( $ n \geq m \geq m' \geq 1 $ ) are coordinates of points in a Gelfand-Zeltin scheme of $\gln$.  
\end{prop}
We give the proof in the Appendix, section B.

{\bf Remark}
If $\alpha >0$ then we can confirm directly that 
the norms of the previous proposition are positive.
The key points are as follows.
For bosonic case, if $s<s'$ then $h^{(s')} - h^{(s)}> p_{s}, \;
\alpha _k ^{(s)} \leq p_{s}$.
For fermionic case, if $s<s'$ then $h^{(s')} - h^{(s)}< -p_{s}, \;
1 \leq m' \leq p_{s}, \; l_{k}^{(s)} \leq p_{s}$.\\

Together with the proposition \ref{p:prop4} and the formula (\ref{eq:boshwnorm}) this proposition gives the norm formulas for the orthogonal eigenbasis of the Spin Calogero-Sutherland Model.

\section{Concluding remarks}

In this paper we have constructed an orthogonal basis of eigenvectors for the Spin Calogero-Sutherland Model and derived product formulas for their norms. Our construction is based on the Gelfand-Zetlin algebra associated with the Yangian symmetry of the Model. It is now natural to ask what are other properties of the eigenvectors described in this paper. What we have in mind is exemplified by the scalar case, where the orthogonal eigenvectors are described by the symmetric Jack polynomials. For the Jack polynomials a number of properties such as triangularity, Cauchy formulas, duality, existence of associated symmetric functions etc.  are known \cite{Macdonaldbook, Stanley}. We believe that  most of these properties have their counterparts for the Calogero-Sutherland Model with spin and plan  to report on this subject in the future.

\vspace{1cm}

\begin{Large} {\bf Appendix } \end{Large}
\appendix
\section{Proof of the theorem \ref{t:FDEC}} 

Recall that for $\mm \in \MC_N$ the subspace $ W^{\mm}_{(-)} \subset \Ncplxn $ was defined in (\ref{eq:W}) as follows
\begin{equation}
W^{\mm}_{(-)}: = \bigcap_{i\: : \:  \mlett_i = \mlett_{i+1}}Ker( P_{i,i+1} + 1), \label{eq:Wdef}
\end{equation}
and that from this definition it follows, in particular, that the dimension of the $W^{\mm}$ is zero unless $ \mm \in \MC^{(n)}_N $ where the set $\MC^{(n)}_N $ is defined in (\ref{eq:nstrict}).

\begin{prop}
If $f \in W^{\mm}_{(-)}$ then $ U^{\mm}_{(-)}f  \in F^{\mm}$ ,\\
and the map $  U^{\mm}_{(-)} : f \longrightarrow  U^{\mm}_{(-)}f $ is an isomorphism of the linear spaces $ W^{\mm}_{(-)}$ and $F^{\mm}$ . \label{p:isomorphism}
\end{prop}
\begin{pf} For an arbitrary $\psi \in E^{\mm}\otimes ( \Ncplxn )$ write
\begin{equation}
\psi = \sum_{\sigma \in S^{\mm}} \Phi_{\s}^{\mm}(z)\otimes
\psi_{\sigma}
\end{equation}
where the components $ \psi_{\sigma} \in \Ncplxn $ are uniquely determined by the $\psi$. We have $ \psi \in F^{\mm}$ if and only if 
\begin{equation}
K_{i,i+1}\psi = - P_{i,i+1}\psi \qquad ( i=1,2,\dots,N-1). \label{eq:P1}
\end{equation}
By virtue of (\ref{eq:KactionPhi1}, \ref{eq:KactionPhi2}) and the linear independence of the $\Phi_{\s}^{\mm}(z)$ the equations (\ref{eq:P1}) are equivalent to  
\begin{eqnarray}
\mbox{for all} & &  \sigma \in S^{\mm} \quad \mbox{and} \quad  i=1,2,\dots,N-1 \nonumber \\
\qquad \quad (P_{i,i+1} + 1) \psi_{\sigma} & = & 0 \quad (\mlett _{\sigma (i)} = \mlett _{\sigma (i+1)} ), \label{eq:P2}\\ 
\psi_{\sigma (i,i+1)} & = & 
\left\{ \begin{array}{ll}
-   \check{R}_{i,i+1} (x) \psi_{\sigma} 
&  (m_{\sigma (i)} > m_{\sigma (i+1)} ), \\
-  \frac{x^2}{x^2 - 1} \check{R}_{i,i+1} (x)\psi_{\sigma} 
&  (m_{\sigma (i)} < m_{\sigma (i+1)} ).
\end{array} \right. 
\; \; 
x := \xi _{i}^{\mm}(\sigma)  - \xi _{i+1}^{\mm}(\sigma) . \label{eq:P3}
\end{eqnarray} 
Notice that the second equation (the case $m_{\sigma (i)} < m_{\sigma (i+1)}$) in (\ref{eq:P3}) is not independent but follows from the first one (the case $m_{\sigma (i)} > m_{\sigma (i+1)}$).

For any $ \sigma \in S^{\mm}$ define a set $ L_{\sigma}$ whose elements are sets $ \{ \psi_{\tau} \}_{\tau \in S^{\mm} } \; (\psi_{\tau} \in \Ncplxn )$. We will say that $ \{ \psi_{\tau} \}_{ \tau \in S^{\mm} } \in L_{\sigma} $ if and only if the following relations are satisfied 
\begin{equation}
(P_{i,i+1} + 1) \psi_{\sigma}  =  0 \quad ( \mbox{for all} \; i \;\mbox{s.t.}\;m _{\sigma (i)} = m_{\sigma (i+1)} ), \label{eq:P4}
\end{equation}
\begin{eqnarray}
\mbox{and for all} & &  \tau \in S^{\mm} \quad \mbox{and} \quad  i=1,2,\dots,N-1 \nonumber \\
\psi_{\tau (i,i+1)} & = & 
\left\{ \begin{array}{ll}
-  \check{R} _{i,i+1} (x) \psi_{\tau} 
&  (\mlett _{\tau (i)} > \mlett _{\tau (i+1)} ); \\
-  \frac{x^2}{x^2-1} \check{R} _{i,i+1} (x)\psi_{\tau} 
&  (\mlett _{\tau (i)} < \mlett _{\tau (i+1)} ),
\end{array} \right. 
\; \; 
x := {\xi _{i}^{\mm}(\tau) }-{
\xi _{i+1}^{\mm}(\tau) }. \label{eq:P5}
\end{eqnarray} 
With this definition we have
\begin{equation}
\psi \in F^{\mm} \Leftrightarrow  (\ref{eq:P2},\ref{eq:P3}) \Leftrightarrow \{ \psi_{\sigma} \}_{\sigma \in S^{\mm} } \in \bigcap_{\sigma \in S^{\mm} }L_{\sigma}. \label{eq:co1}
\end{equation}
\begin{lemma}\begin{equation}
\bigcap_{\sigma \in S^{\mm} }L_{\sigma} = L_{{\id}}.
\end{equation}\end{lemma}

\begin{pf} We will prove that for any $ \sigma \in S^{\mm}$, $ \sigma \prec {\id} $ the inclusion 
\begin{equation}
\{ \psi_{\tau} \}_{\tau \in S^{\mm} } \in \bigcap_{\sigma'\succ\sigma}L_{\sigma'} \label{eq:LL1}
\end{equation}
implies 
\begin{equation}
\{ \psi_{\tau} \}_{\tau \in S^{\mm} } \in L_{\sigma} \label{eq:LL2}.
\end{equation}
Then, since ${\id}$ is the maximal element in $ S^{\mm}$, induction in the order in $ S^{\mm} $ will give   
\begin{equation}
L_{{\id}} \subset \bigcap_{\sigma \in S^{\mm} }L_{\sigma} 
\end{equation}
and the statement of the lemma will follow.

Fix a $\sigma \in S^{\mm}$, $\sigma \prec {\id}$ and assume that (\ref{eq:LL1}) holds. For any $\sigma \in S^{\mm}$, $\sigma \prec {\id}$ there exists $ i\in \{1,2,\dots,N-1\}$ such that $ \mlett_{\sigma(i)} < \mlett_{\sigma(i+1)}$ ( otherwise $\sigma$ must be equal to ${\id}$).

With this $i$ let $\sigma' :=  \sigma (i,i+1)$. Then $ \sigma' \in S^{\mm}$, and by the definition of the ordering in $S^{\mm}$ (\ref{eq:Sord}) we have $\sigma' \succ \sigma$.  

Now take any $j \in \{1,2,\dots,N-1\}$ such that $\mlett_{\sigma(j)} = \mlett_{\sigma(j+1)}$. If such a $j$  does not exist, the implication (\ref{eq:LL1}) $\Rightarrow$ (\ref{eq:LL2}) is obvious. 

The following three situations may take place:
\begin{eqnarray}
& (1) \qquad \qquad  & |j-i| \geq 2 \;, \\
& (2)  \qquad \qquad & j = i+1 \; ,\\
& (3)  \qquad \qquad & j = i-1 \;.
\end{eqnarray}

If (1) holds, then $ \mlett_{\sigma'(j)} =\mlett_{\sigma'(j+1)}$, and by the assumption 
\begin{equation}
\{ \psi_{\tau} \}_{\tau \in S^{\mm} } \in \bigcap_{\sigma'\succ\sigma}L_{\sigma'}  \label{eq:assumption}
\end{equation}
we have   
\begin{equation}
(P_{j,j+1} + 1)\psi_{\sigma'} = 0. 
\end{equation}

The relations (\ref{eq:P5}) give
\begin{equation}
\psi_{\sigma'} = \bar{R}_{i,i+1}(x)\psi_{\sigma}\qquad \left( \bar{R}_{i,i+1}(x) :=-\frac{x^2}{x^2-1}\check{R}_{i,i+1}(x),\; x:= { \xi^{\mm}_{i}(\sigma)}-{\xi^{\mm}_{i+1}(\sigma)} \right). 
\end{equation}
And hence 
\begin{equation}
(P_{j,j+1} + 1)\psi_{\sigma} = 0 
\end{equation}
because $  x:= { \xi^{\mm}_{i}(\sigma)} - {\xi^{\mm}_{i+1}(\sigma)} 
$ = $ \alpha(m_{\sigma(i)}-\mlett_{\sigma(i+1)}) + \sigma(i+1)-\sigma(i) < -1 $  when $ \alpha > 0$ and $ \mlett_{\sigma(i)} < \mlett_{\sigma(i+1)}$ , and therefore the $\bar{R}_{i,i+1}(x)$ is invertible.

Now let the situation  (2) hold (that is $j=i+1$). Then 
\begin{align}
\mlett_{\sigma(i)} < & \mlett_{\sigma(i+1)}  = \mlett_{\sigma(i+2)}, \label{eq:LL3} \\ \sigma(i+2) & =  \sigma(i+1) + 1 . \label{eq:LL4}
\end{align}
Let $\sigma'' := \sigma'(i,i+1) = \sigma (i,i+1)(i+1,i+2).$ We have
\begin{equation}
\begin{array}{lclcl} \sigma''(i) & = & \sigma'(i) & = & \sigma(i+1) \;, \\
\sigma''(i+1) &  = & \sigma'(i+2) &  = & \sigma(i+2) \;,  \\
\sigma''(i+2)&  = & \sigma'(i+1) & = & \sigma(i) \end{array} \label{eq:LL5}
\end{equation}
and hence by (\ref{eq:LL3}) $ \sigma'' \succ \sigma' \succ \sigma $. 

By (\ref{eq:P5}) and (\ref{eq:LL5}) one has  
\begin{equation}
\psi_{\sigma''} = \bar{R}_{i+1,i+2}\left({\xi^{\mm}_{i}(\sigma)}-{\xi^{\mm}_{i+2}(\sigma)}\right)\bar{R}_{i,i+1}\left({\xi^{\mm}_{i}(\sigma)}-{\xi^{\mm}_{i+1}(\sigma)}\right)\psi_{\sigma}. \label{eq:LL6}
\end{equation}
By the assumption (\ref{eq:assumption}) we have 
\begin{equation}
(P_{i,i+1}+ 1)\psi_{\sigma''} = 0. \label{eq:LL7} 
\end{equation}

Since ${\xi^{\mm}_{i+1}(\sigma)}-{\xi^{\mm}_{i+2}(\sigma)} = 1 $, by the Yang-Baxter Equation 
\begin{multline*}
\check{R}_{i,i+1}(\xi_{i+1}^{\mm}(\s) - \xi_{i+2}^{\mm}(\s))\check{R}_{i+1,i+2}(\xi_{i}^{\mm}(\s) - \xi_{i+2}^{\mm}(\s))\check{R}_{i,i+1}(\xi_{i}^{\mm}(\s) - \xi_{i+1}^{\mm}(\s)) = \\ = \check{R}_{i+1,i+2}(\xi_{i}^{\mm}(\s) - \xi_{i+1}^{\mm}(\s))\check{R}_{i,i+1}(\xi_{i}^{\mm}(\s) - \xi_{i+2}^{\mm}(\s))\check{R}_{i+1,i+2}(\xi_{i+1}^{\mm}(\s) - \xi_{i+2}^{\mm}(\s)) 
\end{multline*}
and by  $ \check{R}_{i,i+1}(1) = P_{i,i+1} + 1$ we obtain from (\ref{eq:LL6}, \ref{eq:LL7}) 
\begin{equation}
 \bar{R}_{i+1,i+2}(\xi_{i}^{\mm}(\s) - \xi_{i+1}^{\mm}(\s))\bar{R}_{i,i+1}(\xi_{i}^{\mm}(\s) - \xi_{i+2}^{\mm}(\s))(P_{i+1,i+2} + 1) \psi_{\sigma} = 0.
\end{equation}
Now $( P_{i+1,i+2}+ 1)\psi_{\sigma} = 0$ follows by the invertibility of the operators $\bar{R}_{i+1,i+2}\left({\xi^{\mm}_{i}(\sigma)}-{\xi^{\mm}_{i+1}(\sigma)}\right)$ and $ \bar{R}_{i,i+1}\left({\xi^{\mm}_{i}(\sigma)}-{\xi^{\mm}_{i+2}(\sigma)}\right)$.

The situation (3) is considered in a virtually the same way as the (2) to show that (\ref{eq:assumption}) entails
\begin{equation}
 ( P_{i-1,i} + 1)\psi_{\sigma} = 0
\end{equation}
Thus (\ref{eq:LL1}) implies (\ref{eq:LL2}) and the lemma is proven.\end{pf}

\mbox{} From this lemma and (\ref{eq:co1}) we obtain
\begin{equation}
  \psi \in F^{\mm} \Leftrightarrow \{\psi_{\sigma}\}_{\sigma \in S^{\mm}} \in L_{{\id}}. \label{eq:co2}
\end{equation}

Now we are in position to show that $ v \in W^{\mm}_{(-)}$ implies $ U^{\mm}_{(-)}v \in F^{\mm} $.
Indeed, by the definitions of the $\check{\real}^{(-)}(\sigma)$ (\ref{eq:Rrecursion}) and $W^{\mm}_{(-)}$ (\ref{eq:W}) we have $ \{ \check{\real}^{(-)}(\sigma)v \}_{\sigma \in S^{\mm}} \in L_{{\id}}$ and hence $ U^{\mm}_{(-)}v = \sum_{\sigma \in S^{\mm}} \Phi_{\s}^{\mm}(z)\otimes \check{\real}^{(-)}(\sigma)v $ belongs to $F^{\mm}$ as implied by (\ref{eq:co2}).

Next, we demonstrate surjectivity of the map $U^{\mm}_{(-)}: W^{\mm}_{(-)} \longrightarrow F^{\mm}$. Let  
\begin{equation}
\psi = \sum_{\sigma \in S^{\mm}} \Phi_{\s}^{\mm}(z) \otimes \psi_{\sigma} \in F^{\mm}. 
\end{equation}
Then $\{\psi_{\sigma}\}_{\sigma \in S^{\mm}} \in L_{{\id}} \Rightarrow \psi_{{\id}} \in W^{\mm}_{(-)}$ and solving the relations (\ref{eq:P5}) we find
\begin{equation}
\psi_{\sigma} =  \check{\real}^{(-)}(\sigma) \psi_{id} \qquad \qquad (\sigma \in S^{\mm}).
\end{equation}
Hence $ \psi = U^{\mm}_{(-)}\psi_{{\id}}$ and the surjectivity follows.

Now suppose $ U^{\mm}_{(-)}v = 0 $, $ v \in W^{\mm}_{(-)}$. Due to the linear independence of the non-symmetric Jack polynomials $\Phi_{\s}^{\mm}(z)$ we get  
\begin{equation}
 \check{\real}^{(-)}(\sigma)v = 0 \qquad (\sigma \in S^{\mm}) 
\end{equation}
and in particular $v =0$ which shows injectivity of the map $U^{\mm}_{(-)}$.
This completes the proof of the propositon. \end{pf}

For commuting operators (or complex numbers) $a_i$ $(i=1,2,...,N)$ let $T_0(u)$ be the following monodromy operator
\begin{equation}
T_0(u,\{a_i\}) := L_{0,1}(u,a_1)L_{0,2}(u,a_2) \cdots L_{0,N}(u,a_N).
\end{equation}
Here the $L$-operator is 
\begin{equation}
L_{0,i}(u,a_i) := 1 + \frac{P_{0,i}}{u + a_i}.
\end{equation}

The subspace $F^{\mm}$ for any $\mm \in \MC_N^{(n)}$ is a $Y(\gln)$-module with the action given by the monodromy operator (\ref{eq:That}, \ref{eq:T}) $ \hat{T}_0(u) = T_0(u,\{d_i\})$.  

The space $W^{\mm}_{(-)}$ (\ref{eq:W}) is a  $Y(\gln)$-module as well. Now the action is specified by the monodromy operator $T_0(u,\{\xi^{\mm}_i({\id})\})$. Indeed with this action we have the identity of the Yangian modules (\ref{eq:W=V}): 
\begin{equation}
W^{\mm}_{(-)} =   {V}_{p_1}(h_{\mm}^{(1)})\otimes{V}_{p_2}(h_{\mm}^{(2)})\otimes\ \cdots \otimes {V}_{p_M}(h_{\mm}^{(M)})
\end{equation}
which is established by the standard fusion procedure taking into account that $ \xi^{\mm}_{i+1}({\id}) = 
\xi^{\mm}_{i}({\id}) - 1 $ whenever $ \mlett_i = \mlett_{i+1} $. 

\begin{prop} The map $ U_{(-)}^{\mm}:W^{\mm}_{(-)} \longrightarrow F^{\mm}$ is an intertwiner of the $Y(\gln)$-modules.  \label{p:intertwining}
\end{prop}

\begin{pf}
The intertwining property of the $R$-matrix: 
\begin{multline}
L_{0,i}(u,\xi_i^{\mm}(\s))L_{0,i+1}(u,\xi_{i+1}^{\mm}(\s))\check{R}_{i,i+1}(\xi_i^{\mm}(\s(i,i+1))-\xi_{i+1}^{\mm}(\s(i,i+1)) = \\ 
= \check{R}_{i,i+1}(\xi_i^{\mm}(\s(i,i+1))-\xi_{i+1}^{\mm}(\s(i,i+1))L_{0,i}(u,\xi_i^{\mm}(\s(i,i+1)))L_{0,i+1}(u,\xi_{i+1}^{\mm}(\s(i,i+1))),
\end{multline}
and (\ref{eq:d-eigenvectors}) entail the following chain of equations $(v \in W^{\mm}_{(-)})$: 
\begin{multline}
 \hat{T}_{0}(u)U_{(-)}^{\mm }v =\sum_{{\sigma } \in S^{\mm }}
 \Phi_{\sigma }^{\mm}(z)\otimes  
T_{0} ( u ; \{ \xi _{i}^{\mm}(\sigma) \} ) \check{\real}^{(-)}(\sigma)v = \\
= \sum_{{\sigma } \in S^{\mm }}
 \Phi_{\sigma }^{\mm}(z)\otimes \check{\real}^{(-)}(\sigma)  
T_{0} ( u , \{ \xi _{i}^{\mm }({\id} )  \} )v =  U_{(-)}^{\mm }T_{0} ( u , \{ \xi _{i}^{\mm }({\id} ) \})v
\end{multline}
\end{pf}

The propositions \ref{p:isomorphism} and \ref{p:intertwining} imply the statement of the theorem \ref{t:FDEC}.

\section{Proof of the proposition \ref{norm}}


Let us  define the vector 
\begin{equation}
\bar{v}^{(\kappa)}_{ \Lambda ^{(1)}, \dots ,\Lambda ^{(M)}} =
U_{(\kappa)}^{\mm } \cdot \left( \prod_{(s,t) \atop{1 \leq t \leq \l ^{(s)}_{n,m'}- 
\l ^{(s)}_{m,m'}}} b_m(\nu_{m,m'}^{(s)}-t)
\right) \cdot (U_{(\kappa)}^{\mm})^{-1} \Omega_{\mm}^{(\kappa)}, \; \; \; \; (\kappa = -,+),
\end{equation}
where $b_{m}(u)$ and $\nu _{m,m'}^{(s)}-t$ are defined in (\ref{bm}), (\ref{nu}). 
 Notice that the following relations are satisfied:
\begin{equation}
v^{(\kappa)}_{ \Lambda ^{(1)}, \dots ,\Lambda ^{(M)}}=
f_{\kappa}( \Lambda ^{(1)}, \dots ,\Lambda ^{(M)})
\bar{v}^{(\kappa)}_{ \Lambda ^{(1)}, \dots ,\Lambda ^{(M)}} \; \; 
\mbox{ for some scalar function } f_{\kappa}( \cdot ). \label{f}
\end{equation}
The calculation of the function $f_{\kappa}( \cdot )$ can be done by comparing
 the ratio of $b_m(u )$ and $b^{(\kappa)}_m(u ) \; (\kappa = - , + )$.  
We will calculate the norms of
 $\bar{v}^{(\kappa)}_{ \Lambda ^{(1)}, \dots ,\Lambda ^{(M)}}$.
Then we will get the norms of 
$v^{(\kappa)}_{ \Lambda ^{(1)}, \dots ,\Lambda ^{(M)}}$.

To calculate the norms of $\bar{v}^{(\kappa)}_{ \Lambda ^{(1)}, \dots
 ,\Lambda ^{(M)}}$, we will derive  recursion relations between
\begin{equation*}
( \Lambda ^{(1)}, \dots ,\Lambda ^{(i)}, \dots ,\Lambda ^{(M)}) \quad \text{and}\quad ( \Lambda ^{(1)}, \dots ,\Lambda ^{(i)}+ e_{m,m'},
 \dots ,\Lambda ^{(M)})
\end{equation*}
 and will solve them. 
Here $\Lambda ^{(i)}+ e_{m,m'} $ is the Gelfand-Zetlin scheme,
 whose $(j,j')$--elements are
$\l _{j,j'} + \delta _{j,m} \delta _{j',m'}$.

\begin{prop} \label{rec}
\begin{equation}
\langle  \bar{v}^{(\kappa)}_{ \Lambda ^{(1)}, \dots ,\Lambda ^{(i)},
 \dots ,\Lambda ^{(M)}},
\bar{v}^{(\kappa)}_{ \Lambda ^{(1)}, \dots ,\Lambda ^{(i)}, \dots ,
\Lambda ^{(M)}} \rangle_{(\kappa)}
= 
\langle  \bar{v}^{(\kappa)}_{ \Lambda ^{(1)}, \dots ,\Lambda ^{(i)}+ e_{m,m'},
 \dots ,\Lambda ^{(M)}},
\bar{v}^{(\kappa)}_{ \Lambda ^{(1)}, \dots ,\Lambda ^{(i)}+ e_{m,m'},
 \dots ,\Lambda ^{(M)}} \rangle_{(\kappa)} \cdot
\end{equation}
\begin{equation}
\cdot
\varpi _{m+1,+}(\nu )
\varpi _{m-1,+}(\nu -1)
\varpi _{m,+}(\nu )^{-1}
\bar{\varpi }_{m,+}(\nu -1). \nonumber
\end{equation}
Here $\nu = m'-\l _{m,m'}^{(s)}-1-h_{\mm}^{(s)}  $,
 and $ \varpi _{k,+}(u ) , \; \bar{\varpi }_{k,+}(u)$
 are defined by the following relations:
\begin{equation}
\varpi _{k,+}(u ) \bar{v}^{(\kappa)}_{ \Lambda ^{(1)}, \dots ,
\Lambda ^{(i)}+ e_{m,m'}, \dots ,\Lambda ^{(M)}} =
a _{k}(u) \bar{v}^{(\kappa)}_{ \Lambda ^{(1)}, \dots ,
\Lambda ^{(i)}+ e_{m,m'}, \dots ,\Lambda ^{(M)}}.
\end{equation}
\begin{equation}
\bar{\varpi }_{k,+}(u)= \lim _{u' \rightarrow u}
(u-u')\varpi _{k,+}(u' ).
\end{equation}
\end{prop}
{ Proof. }
  
By using the relation (\ref{con}) and (\ref{f}), we get
\begin{equation}
\langle  \bar{v}^{(\kappa)}_{ \Lambda ^{(1)}, \dots ,\Lambda ^{(i)},
 \dots ,\Lambda ^{(M)}},
\bar{v}^{(\kappa)}_{ \Lambda ^{(1)}, \dots ,\Lambda ^{(i)}, \dots ,
\Lambda ^{(M)}} \rangle_{(\kappa)} =
\end{equation}
\begin{equation}
= 
\langle  \bar{v}^{(\kappa)}_{ \Lambda ^{(1)}, \dots ,\Lambda ^{(i)}+ e_{m,m'},
 \dots ,\Lambda ^{(M)}}, U_{(\kappa)}^{\mm } \cdot c_m(\nu -1) b_m(\nu -1)
\cdot (U_{(\kappa)}^{\mm})^{-1}
\bar{v}^{(\kappa)}_{ \Lambda ^{(1)}, \dots ,\Lambda ^{(i)}+ e_{m,m'},
 \dots ,\Lambda ^{(M)}} \rangle_{(\kappa)} \nonumber
\end{equation} 
\begin{equation}
= \lim _{\nu ' \rightarrow \nu }
\langle  \bar{v}^{(\kappa)}_{ \Lambda ^{(1)}, \dots ,\Lambda ^{(i)}+ e_{m,m'},
 \dots ,\Lambda ^{(M)}}, U_{(\kappa)}^{\mm } \cdot c_m(\nu -1) b_m(\nu '-1)
\cdot (U_{(\kappa)}^{\mm})^{-1}
\bar{v}^{(\kappa)}_{ \Lambda ^{(1)}, \dots ,\Lambda ^{(i)}+ e_{m,m'},
 \dots ,\Lambda ^{(M)}} \rangle_{(\kappa)} . \nonumber
\end{equation}
On the other hand,  the relation (\ref{cb}) gives  
\begin{multline}
 c_m(\nu -1) b_m(\nu '-1) =b_m(\nu '-1) c_m(\nu -1)+ \\
+ \frac{1}{\nu - \nu '} \{ d_m(\nu -1) a_m(\nu '-1) - d_m(\nu' -1) a_m(\nu -1)\}.
\label{cbbc}
\end{multline}
Since $a_m(\nu -1) \cdot (U_{(\kappa)}^{\mm})^{-1}
\bar{v}^{(\kappa)}_{ \Lambda ^{(1)}, \dots ,\Lambda ^{(i)}+ e_{m,m'},
 \dots ,\Lambda ^{(M)}} =0$, and 
\begin{equation} \lim _{\nu ' \rightarrow \nu} 
\frac{1}{\nu - \nu '}a_m(\nu '-1)
\bar{v}^{(\kappa)}_{ \Lambda ^{(1)}, \dots ,\Lambda ^{(i)}+ e_{m,m'},
 \dots ,\Lambda ^{(M)}} = \bar{\varpi }_{m,+}(\nu -1)
\bar{v}^{(\kappa)}_{ \Lambda ^{(1)}, \dots ,\Lambda ^{(i)}+ e_{m,m'},
 \dots ,\Lambda ^{(M)}},  \nonumber 
\end{equation}
we have
\begin{equation}
 \langle  \bar{v}^{(\kappa)}_{ \Lambda ^{(1)}, \dots ,\Lambda ^{(i)},
 \dots ,\Lambda ^{(M)}},
\bar{v}^{(\kappa)}_{ \Lambda ^{(1)}, \dots ,\Lambda ^{(i)}, \dots ,
\Lambda ^{(M)}} \rangle_{(\kappa)} =
\label{d}
\end{equation}
\begin{equation}
= \langle  \bar{v}^{(\kappa)}_{ \Lambda ^{(1)}, \dots ,\Lambda ^{(i)}+ e_{m,m'},
 \dots ,\Lambda ^{(M)}}, U_{(\kappa)}^{\mm } \cdot b_m(\nu -1) c_m(\nu -1)
\cdot (U_{(\kappa)}^{\mm})^{-1} \bar{v}^{(\kappa)}_{ \Lambda ^{(1)},
 \dots ,\Lambda ^{(i)}+ e_{m,m'},
 \dots ,\Lambda ^{(M)}} \rangle_{(\kappa)} + \nonumber
\end{equation}
\begin{equation} 
+ \langle  \bar{v}^{(\kappa)}_{ \Lambda ^{(1)}, \dots ,\Lambda ^{(i)}+ e_{m,m'},
 \dots ,\Lambda ^{(M)}}, U_{(\kappa)}^{\mm } \cdot d_m(\nu -1) 
\cdot (U_{(\kappa)}^{\mm})^{-1} \bar{v}^{(\kappa)}_{ \Lambda ^{(1)},
 \dots ,\Lambda ^{(i)}+ e_{m,m'},
 \dots ,\Lambda ^{(M)}} \rangle_{(\kappa)} \cdot  \bar{\varpi }_{m,+}(\nu -1)= \nonumber
\end{equation}
\begin{equation}
  = \langle  \bar{v}^{(\kappa)}_{ \Lambda ^{(1)}, \dots ,\Lambda ^{(i)}+ e_{m,m'},
 \dots ,\Lambda ^{(M)}}, U_{(\kappa)}^{\mm } \cdot b_m(\nu -1) c_m(\nu -1)
\cdot (U_{(\kappa)}^{\mm})^{-1} \bar{v}^{(\kappa)}_{ \Lambda ^{(1)},
 \dots ,\Lambda ^{(i)}+ e_{m,m'},
 \dots ,\Lambda ^{(M)}} \rangle_{(\kappa)} + 
\nonumber 
\end{equation}
\begin{multline}
+ \langle  \bar{v}^{(\kappa)}_{ \Lambda ^{(1)}, \dots ,\Lambda ^{(i)}+ e_{m,m'},
 \dots ,\Lambda ^{(M)}}, U_{(\kappa)}^{\mm } \cdot b_m(\nu -1) c_m(\nu ) 
\cdot (U_{(\kappa)}^{\mm})^{-1} \bar{v}^{(\kappa)}_{ \Lambda ^{(1)},
 \dots ,\Lambda ^{(i)}+ e_{m,m'},
 \dots ,\Lambda ^{(M)}} \rangle_{(\kappa)} \cdot \\ \cdot \bar{\varpi }_{m,+}(\nu -1)
\varpi _{m,+}(\nu )^{-1}
\nonumber 
\end{multline}
\begin{multline}
+
 \langle  \bar{v}^{(\kappa)}_{ \Lambda ^{(1)}, \dots ,\Lambda ^{(i)}+ e_{m,m'},
 \dots ,\Lambda ^{(M)}}, \bar{v}^{(\kappa)}_{ \Lambda ^{(1)},
 \dots ,\Lambda ^{(i)}+ e_{m,m'},
 \dots ,\Lambda ^{(M)}} \rangle_{(\kappa)} 
\cdot \\ \cdot \bar{\varpi }_{m,+}(\nu -1)
\varpi _{m,+}(\nu )^{-1} \varpi _{m+1,+}(\nu ) \varpi _{m-1,+}(\nu -1). 
\nonumber
\end{multline}
 In (\ref{d}), we used the relation (\ref{daaa}).
 Then if we show the following lemma, we have proved
 the Proposition \ref{rec}.  \hfill\halmos

\begin{lemma} \label{lem}
In the situation of the Proposition \ref{rec}, we have
\begin{equation} 
\langle  \bar{v}^{(\kappa)}_{ \Lambda ^{(1)}, \dots ,\Lambda ^{(i)}+ e_{m,m'},
 \dots ,\Lambda ^{(M)}}, U_{(\kappa)}^{\mm } \cdot b_m(\nu -1) c_m(\nu -1)
\cdot (U_{(\kappa)}^{\mm})^{-1} \bar{v}^{(\kappa)}_{ \Lambda ^{(1)},
 \dots ,\Lambda ^{(i)}+ e_{m,m'},
 \dots ,\Lambda ^{(M)}} \rangle_{(\kappa)} .
\label{eqlem}
\end{equation}
\begin{equation}
+ \langle  \bar{v}^{(\kappa)}_{ \Lambda ^{(1)}, \dots ,\Lambda ^{(i)}+ e_{m,m'},
 \dots ,\Lambda ^{(M)}}, U_{(\kappa)}^{\mm } \cdot b_m(\nu -1) c_m(\nu ) 
\cdot (U_{(\kappa)}^{\mm})^{-1} \bar{v}^{(\kappa)}_{ \Lambda ^{(1)},
 \dots ,\Lambda ^{(i)}+ e_{m,m'},
 \dots ,\Lambda ^{(M)}} \rangle_{(\kappa)} 
\nonumber 
\end{equation}
\begin{equation}
\cdot  \bar{\varpi }_{m,+}(\nu -1)
\varpi _{m,+}(\nu )^{-1}=0 \nonumber
\end{equation}
\end{lemma}
{ Proof.}

By using the relation (\ref{cb}), we can show inductively that if
$\Lambda ^{(i)}+ t \, e_{m,m'} \in {\cal S}_{\l ^{(r)}}$ then
\begin{equation}
l.h.s.\; \;  of \; \; (\mbox{\ref{eqlem}}) = 
\langle  \bar{v}^{(\kappa)}_{ \Lambda ^{(1)}, \dots ,\Lambda ^{(i)}+ e_{m,m'},
 \dots ,\Lambda ^{(M)}},
\end{equation}
\begin{equation}
U_{(\kappa)}^{\mm } \cdot b_m(\nu -1) \cdots b_m(\nu -t)
c_m(\nu -1)
\cdot (U_{(\kappa)}^{\mm})^{-1} \bar{v}^{(\kappa)}_{ \Lambda ^{(1)},
 \dots ,\Lambda ^{(i)}+ t \, e_{m,m'},
 \dots ,\Lambda ^{(M)}} + 
\nonumber 
\end{equation}
\begin{equation}
 U_{(\kappa)}^{\mm } \cdot b_m(\nu -1) \cdots b_m(\nu -t)
c_m(\nu )
\cdot (U_{(\kappa)}^{\mm})^{-1} \bar{v}^{(\kappa)}_{ \Lambda ^{(1)},
 \dots ,\Lambda ^{(i)}+ t \, e_{m,m'},
 \dots ,\Lambda ^{(M)}} \cdot  \bar{\varpi }_{m,+}(\nu -1)
\varpi _{m,+}(\nu )^{-1} \rangle_{(\kappa)} . \nonumber
\end{equation}
Let $t$ be the maximal number s.t. 
$\Lambda ^{(i)}+ t \, e_{m,m'} \in {\cal S}_{\l ^{(r)}}$.
\mbox{} From the relations (\ref{adj}), we get
\begin{equation}
l.h.s.\; \;  of \; \; (\mbox{\ref{eqlem}}) 
\end{equation}
\begin{equation}
= \langle U_{(\kappa)}^{\mm} \cdot c_m(\nu -t) \cdots c_m(\nu -1)(U_{(\kappa)}^{\mm})^{-1}
\bar{v}^{(\kappa)}_{ \Lambda ^{(1)}, \dots ,\Lambda ^{(i)}+ e_{m,m'},
 \dots ,\Lambda ^{(M)}},
\end{equation}
\begin{equation}
 U_{(\kappa)}^{\mm } \cdot c_m(\nu -1)
\cdot (U_{(\kappa)}^{\mm})^{-1} \bar{v}^{(\kappa)}_{ \Lambda ^{(1)},
 \dots ,\Lambda ^{(i)}+ t \, e_{m,m'},
 \dots ,\Lambda ^{(M)}} + 
\nonumber 
\end{equation}
\begin{equation}
+  U_{(\kappa)}^{\mm } \cdot c_m(\nu )
\cdot (U_{(\kappa)}^{\mm})^{-1} \bar{v}^{(\kappa)}_{ \Lambda ^{(1)},
 \dots ,\Lambda ^{(i)}+ t \, e_{m,m'},
 \dots ,\Lambda ^{(M)}} \cdot  \bar{\varpi }_{m,+}(\nu -1)
\varpi _{m,+}(\nu )^{-1} \rangle_{(\kappa)} . \nonumber
\end{equation}
If we apply repeatedly the Theorem 3.5. written in the paper \cite{nt2}
(here $\gamma _{m,m'}^{(i)}$ is some constant),
\begin{equation}
U_{(\kappa)}^{\mm} c_m (\nu_{m,m'}^{(i)} ) (U_{(\kappa)}^{\mm})^{-1}
 \bar{v}^{(\kappa)}_{ \Lambda ^{(1)}, \dots ,\Lambda ^{(i)},
 \dots ,\Lambda ^{(M)}} = \left\{
\begin{array}{cl}
\gamma _{m,m'}^{(i)}  \bar{v}^{(\kappa)}_{ \Lambda ^{(1)},
 \dots ,\Lambda ^{(i)} + e_{m,m'},
 \dots ,\Lambda ^{(M)}} & \mbox{ if }
\Lambda ^{(i)}+ e_{m,m'} \in {\cal S}_{\l ^{(r)}}; \\
0 & \mbox{ otherwise,}
\end{array}
\right.
\end{equation}
we have
\begin{equation}
U_{(\kappa)}^{\mm} \cdot c_m(\nu -t) \cdots c_m(\nu -1)(U_{(\kappa)}^{\mm})^{-1}
\bar{v}^{(\kappa)}_{ \Lambda ^{(1)}, \dots ,\Lambda ^{(i)}+ e_{m,m'},
 \dots ,\Lambda ^{(M)}}=0.
\end{equation}
So we get the Lemma \ref{lem}. \hfill\halmos

\vspace{.2in}
Let $\backslash \kappa_{m,m'} / $ be the Gelfand Zetlin scheme which
 corresponds to the highest weight vector, the highest weights are
 $(\l_{n,1} , \dots , \l_{n,n}) $. (i.e. $\kappa_{m,m'}=\l _{n,m'}$
 for all possible $m,m'$.) If we solve the recursive relations of the 
 Proposition \ref{rec}, we get

\begin{prop} \label{fo}
\begin{equation}
\langle  \bar{v}^{(\kappa)}_{ \Lambda ^{(1)}, \dots ,\Lambda ^{(i)},
 \dots ,\Lambda ^{(M)}},
\bar{v}^{(\kappa)}_{ \Lambda ^{(1)}, \dots ,\Lambda ^{(i)}, \dots ,
\Lambda ^{(M)}} \rangle_{(\kappa)}
= 
\langle \bar{v}^{(\kappa)}_{h.w.v. }  , \bar{v} ^{(\kappa)}_{h.w.v. } \rangle_{(\kappa)} \cdot
\prod _{(m,m') \atop{m' \leq m}} 
\left\{
\prod _{(s,s') \atop{\mbox{{\scriptsize for all pairs}} }} 
\left\{ 
\prod _{a=\lambda _{m,m'}^{(s)}}^{\kappa _{m,m'}^{(s)} -1}
\left\{
(-1)^{\delta _{s,s'}}
\right. \right. \right.
\nonumber
\end{equation}
\begin{equation}
\prod _{j=1}^{m'}(m'-j-a+ \kappa _{m+1,j}^{(s')} +h_{\mm}^{(s')}-h_{\mm}^{(s)} )
\prod _{j=m'+1}^{m+1} (m'-j-a+ \lambda _{m+1,j}^{(s')} +h_{\mm}^{(s')}-h_{\mm}^{(s)} )
\nonumber
\end{equation}
\begin{equation}
\left.
\prod _{j=1}^{m'-1} (m'-j-1-a + \kappa _{m-1,j}^{(s')} +h_{\mm}^{(s')}-h_{\mm}^{(s)} )
\prod _{j=m'}^{m-1} (m'-j-1-a + \lambda _{m-1,j}^{(s')} +h_{\mm}^{(s')}-h_{\mm}^{(s)} )
\right\} 
\nonumber
\end{equation}
\begin{equation}
\left.
\prod _{j=1}^{m'-1} 
\frac{(m'-j- \kappa _{m,m'}^{(s)} + \kappa _{m,j}^{(s')} +h_{\mm}^{(s')}-h_{\mm}^{(s)} )}{
(m'-j- \lambda _{m,m'}^{(s)}+ \kappa _{m,j}^{(s')}+h_{\mm}^{(s')}-h_{\mm}^{(s)} )}
\prod _{j=m'+1}^{m}
\frac{(m'-j- \kappa _{m,m'}^{(s)} + \lambda _{m,j}^{(s')}+h_{\mm}^{(s')}-h_{\mm}^{(s)})}{
(m'-j- \lambda _{m,m'}^{(s)} + \lambda _{m,j}^{(s')} +h_{\mm}^{(s')}-h_{\mm}^{(s)} )}
\right\}
\nonumber
\end{equation}
\begin{equation}
\prod _{(s,s') \atop{s < s'} } 
\left. 
\frac{(- \kappa _{m,m'}^{(s)} + \kappa _{m,m'}^{(s')} +h_{\mm}^{(s')}-h_{\mm}^{(s)} )}{
( - \lambda _{m,m'}^{(s)} + \lambda _{m,m'}^{(s')} +h_{\mm}^{(s')}-h_{\mm}^{(s)} )}
\right\}
\nonumber
\end{equation}
\end{prop}

If we rewrite Proposition \ref{fo} for Bosonic (resp. Fermionic) case 
and take into account the function $f_{\kappa} (\cdot )$, we get the Proposition \ref{norm}.

\end{document}